\documentclass[sigconf,authorversion]{acmart}

\usepackage{booktabs}

\usepackage[inline]{enumitem}

\setenumerate{label=(\roman*), leftmargin=2em}

\newcommand{\inlineheading}[1]{\vspace{0.5em}\noindent\textbf{{#1}.}}

\newcommand{\boldparagraph}[1]{\vspace{0.5em}\noindent\textbf{{#1}.}}

\usepackage{wasysym}
\newcommand{\cmark}{\CIRCLE}%
\newcommand{\xmark}{\Circle}

\usepackage{fontawesome5}
\newcommand{\macos}{\faDesktop}
\newcommand{\android}{\faAndroid}
\newcommand{\ios}{\faApple}

\usepackage{listings} 
\copyrightyear{2024}
\acmYear{2024}
\setcopyright{acmlicensed}\acmConference[ARES 2024]{The 19th International Conference on Availability, Reliability and Security}{July 30-August 2, 2024}{Vienna, Austria}
\acmBooktitle{The 19th International Conference on Availability, Reliability and Security (ARES 2024), July 30-August 2, 2024, Vienna, Austria}
\acmDOI{10.1145/3664476.3664478}
\acmISBN{979-8-4007-1718-5/24/07}

\usepackage{pdfprivacy}

\begin{document}

\title{A Privacy Measure Turned Upside Down? Investigating the Use of HTTP Client Hints on the Web}

\author{Stephan Wiefling}
\authornote{Stephan Wiefling started this research in March 2022 while working at H-BRS University of Applied Sciences. He was not involved in this project between July 25th and October 14th, 2022, where he worked at a Big Tech company. All opinions expressed are his own and not necessarily those of his current or former employers.}
\email{clienthints@swiefling.de}
\orcid{0000-0001-7917-6065}
\affiliation{%
    	\institution{swiefling.de \&\\H-BRS University of Applied Sciences}
    	\city{Sankt Augustin}
    	\country{Germany}
}

\author{Marian Hönscheid}
\email{marian.hoenscheid@smail.inf.h-brs.de}
\orcid{0009-0004-3196-1549}
\affiliation{%
	\institution{H-BRS University of Applied Sciences}
	\city{Sankt Augustin}
	\country{Germany}
}

\author{Luigi Lo Iacono}
\email{luigi.lo_iacono@h-brs.de}
\orcid{0000-0002-7863-0622}
\affiliation{%
	\institution{H-BRS University of Applied Sciences}
	\city{Sankt Augustin}
	\country{Germany}
}

\renewcommand{\shortauthors}{Wiefling et al.}

\begin{abstract}

HTTP client hints are a set of standardized HTTP request headers designed to modernize and potentially replace the traditional user agent string. While the user agent string exposes a wide range of information about the client's browser and device, client hints provide a controlled and structured approach for clients to selectively disclose their capabilities and preferences to servers. Essentially, client hints aim at more effective and privacy-friendly disclosure of browser or client properties than the user agent string.

We present a first long-term study of the use of HTTP client hints in the wild. We found that despite being implemented in almost all web browsers, server-side usage of client hints remains generally low. However, in the context of third-party websites, which are often linked to trackers, the adoption rate is significantly higher. This is concerning because client hints allow the retrieval of more data from the client than the user agent string provides, and there are currently no mechanisms for users to detect or control this potential data leakage. Our work provides valuable insights for web users, browser vendors, and researchers by exposing potential privacy violations via client hints and providing help in developing remediation strategies as well as further research. \end{abstract}



\maketitle

\section{Introduction}
\label{sec:introduction}
The user agent string (UAS)~\cite{nielsen_http_1996} has long played an important role for websites to obtain information about the client's browser, operating system (OS) and device information. In web browsers, it is commonly a string with the format \texttt{Mozilla/5.0 (<system-informa\-tion>) <platform> (<platform-details>) <extensions>}~\cite{mdn_user_2024} (see Table~\ref{tab:example-uas}). Websites use the UAS for security purposes to detect bots~\cite{vastel_fp-crawlers_2020} or legitimate users~\cite{Wiefling_Pump_2022}, but also in a privacy-invasive way to track users across websites~\cite{eckersley_how_2010,pugliese_long-term_2020}. As a countermeasure, initiatives by major web browser vendors recently started to deprecate the static and information-rich UAS and replacing it with Hypertext Transfer Protocol (HTTP) client hints (CHs)~\cite{ney_prepare_2022,weiss_client_2023,chromium_user_2023}. These aim to increase user privacy by default, submitting a low-entropy version of the UAS. To obtain high-entropy features---which are the default in the UAS---websites need to explicitly request them from a user's web browser via a dedicated HTTP response header. Users will not directly notice this interaction, unless they inspect the HTTP responses in the browser's developer tools. Still, this proactive approach reveals the type of information that websites aim to collect about users.

\begin{table}[t]
	\vspace{1.7em}
    \centering
    \caption{Example UASs for different web browsers, devices, and OSs}
    \resizebox{\linewidth}{!}{
    \begin{tabular}{@{}p{1.4\linewidth}@{}}
        \toprule
        \textbf{Chrome 102 on Desktop PC with Windows 10} \\
        \texttt{Mozilla/5.0 (Windows NT 10.0; WOW64) AppleWebKit/537.36 (KHTML, like Gecko) Chrome/102.0.5005.63 Safari/537.36}\\[0.5em]
        \textbf{Firefox 117 on iPhone with iOS 16}\\
        \texttt{Mozilla/5.0 (iPhone; CPU iPhone OS 16\_6 like Mac OS X) AppleWebKit/605.1.15 (KHTML, like Gecko) FxiOS/117.2 Mobile/15E148 Safari/605.1.15}\\
        \bottomrule
    \end{tabular}%
    }
    \label{tab:example-uas}
\end{table}

Apart from these essentially privacy-respecting innovations, however, HTTP CHs can provide much more information about the client than the UAS when explicitly requested by the server%
. This means that they can potentially be used to track users even better than with the UAS. It is therefore very important to understand whether HTTP CHs are misused for identification and tracking purposes in practice. If this is done without users' consent, which would be the case when a user visits a website for the first time without sending active consent, this could potentially violate current privacy laws like the GDPR~\cite{european_union_gdpr_2016} and CCPA~\cite{california_ccpa_2018}.

Besides that, there are also legitimate and legal purposes for requesting HTTP CH data, like for security measures to prevent unauthorized access~\cite{wiefling_privacy_2021}. Very relevant here is the case of risk-based authentication (RBA)~\cite{Wiefling_Pump_2022,Wiefling_Whats_2021,makowski_evaluation_2023,wiefling_is_2019,doerfler_evaluating_2019,Unsel_Risk_2023}, which is recommended by national authorities~\cite{biden_jr_executive_2021,grassi_digital_2017,national_cyber_security_centre_cloud_2018,australian_australian_2021} and often uses the UAS alongside the IP address to determine in a risk estimation whether the legitimate user is signing in. Since the UAS provides less information with its deprecation, we hypothesize that RBA-using websites might adopt HTTP CHs instead. As a major difference, the HTTP CH requests can now give indications of potential features used for the risk estimation, which was not possible with the UAS before. Learning this from other online services can help RBA system designers to focus on effective features to protect their users from attacks, such as credential stuffing~\cite{akamai_loyalty_2020} and password spraying~\cite{haber_attack_2020}.

\boldparagraph{Research Questions}
\label{sec:research-questions}
HTTP CHs and their effects on online services and users in the Web have not been studied intensively in literature before. Understanding them is, however, very important, as their usage might have consequences for online users' privacy and security. To close this research gap, we formulated the following research questions.

\newlist{RQLIST}{enumerate}{1}
\setlist[RQLIST]{label=\bfseries RQ\arabic*:, leftmargin=2.7em, parsep=0em}

\newlist{RQ2LIST}{enumerate}{2}
\setlist[RQ2LIST]{label=\bgroup\bfseries \alph*)\egroup,leftmargin=1.4em, parsep=0em}

\begin{RQLIST}
    \item \textbf{Adoption of HTTP CHs} \begin{RQ2LIST}
        \item How do website start pages adjust to the transition from UAS to HTTP CHs?
        \item Do websites request different HTTP CHs on the start page and on the login page?
        \item How do login pages using RBA adopt to the transition from UAS to HTTP CHs?
        \item How wide-spread are HTTP CH requests on login pages in practice, also regarding requests from embedded third party domains and known web trackers?
    \end{RQ2LIST}
    \item \textbf{Requested HTTP CHs} \begin{RQ2LIST}
        \item What HTTP CH data do websites and known web trackers request on login pages%
        ?
        \item What HTTP CH data do websites request on their login page when using RBA?
        \item Do geographical location and the Internet Service Pro\-vider (ISP) make a difference in how websites request HTTP CH data?
        \item What HTTP CHs do different categories of websites request and to what level of detail?
        \item What HTTP CHs do different RBA-instrumented websites request and to what level of detail?
    \end{RQ2LIST}
    \item \textbf{Impact} \begin{RQ2LIST}
        \item How interconnected are HTTP CH requests by third party domains with different login pages?
        \item How much information do web browsers provide when receiving HTTP CH requests?
        \end{RQ2LIST}
\end{RQLIST}

\boldparagraph{Contributions}
\label{sec:contributions}
By answering the research questions, we contribute the following to the current body of knowledge:

\begin{itemize}
    \item \textbf{First historical overview of HTTP CH usage on the Web}:
    Using historical crawling data, we provide an overview of how the adoption of HTTP CHs on websites of the full Tranco list\footnote{Available at \url{https://tranco-list.eu/list/GZP2K}}  with $\approx$8M of the most popular websites worldwide (Tranco 8M) increased since its first appearance in July 2017. Combined with our own crawling data of 327K identified login pages on the Tranco 8M Uniform Resource Locators (URLs), we also show that websites tend to request more high-entropy HTTP CHs on login pages compared to their corresponding start page.
    \item \textbf{First large-scale analysis of HTTP CH usage on login pages worldwide}:
    Using our own crawling data, we show the type and amount of user information that 327K login pages request with HTTP CHs. We identify differences in requested level of detail across different website categories. We also show that websites known to use some sort of RBA request higher level of detail than those without.
    \item \textbf{HTTP CH analysis on RBA-using websites}:
    We show the type of HTTP CH features that RBA-using websites requested and potentially used for their RBA risk estimation.
    \item \textbf{Influence of third parties, trackers, regions and ISPs}:
    We provide an overview of how third parties and known trackers request HTTP CH data, and how they are connected to other websites. We show that third parties request a high to very high level of detail from the user. We also show that they are interconnected with 18.1\% of the crawled 1,000 most popular (Tranco 1K) and 13.8\% of the 5,000 most popular (Tranco 5K) websites, allowing user profiling across popular websites. Our results also show that geographical region and ISP have an effect on whether websites request HTTP CH data. 
    \item \textbf{Open Data}:
    We provide the HTTP CH results from the crawled login pages as an open data set\footnote{Available at \url{https://github.com/das-group/http-client-hints-dataset}}
    . This can be used to reproduce our results and to conduct own research on HTTP CHs. For researchers following ethical standards, we also provide our data set of login page URLs on request\footnote{We decided not to publish this data openly to limit potential misuse by attackers (e.g., automated credential stuffing attacks), although it could be possible that smart attackers already own such URL data themselves.}.
\end{itemize}

Our findings support developers, and security and privacy engineers to gain insights on how HTTP CHs are used in practice and how they are used in relation to tracking and RBA. Researchers also get an overview of the popularity of HTTP CHs on the Web, and obtain an open data set to do more research on HTTP CHs in practice. Privacy experts and policy makers also gather information on whether HTTP CHs might be used to breach users' privacy, although they were intended and introduced as a privacy measure.

\section{HTTP Client Hints (HTTP CHs)}
\label{sec:client-hints}

Before introducing the study, we first give an overview of HTTP CHs and their implementation in web browsers.

\subsection{Background}
HTTP CHs~\cite{weiss_client_2023} aim to be a privacy-preserving measure to request client information from a user's web browser. In contrast to the UAS, a web browser only provides more fine-grained browser, OS, and device information when a server sent an \texttt{Accept-CH} HTTP response header in a previous protocol message exchange. This header contains the type of information that the client should provide~\cite{weiss_client_2023} (see Table~\ref{tab:client-hint-classification} for the list of all possible HTTP CHs). The client will cache this header and remember its values. Therefore, in all subsequent client requests, the client's browser will always provide the requested information (e.g., the fine-grained browser and OS version). Note that by means of this behavior, the website can recognize whether a particular user has already visited the website in the past. Beyond that, servers can also request device and network information that was not included in the UAS, like the client-measured round-trip time (RTT) or the display resolution. Note that some of this information could also be accessed via JavaScript functions. However, HTTP CHs do not require JavaScript, so preventing their data submission via deactivated JavaScript functionality is no longer possible.

With the roll out of HTTP CHs in web browsers, Chromium-based browsers started to deprecate the UAS and changed it to a low entropy version that makes it less distinguishable from other users~\cite{chromium_user_2023}. Therefore, to obtain higher entropy information, servers have to request it via HTTP CHs. 

The first draft of the idea was published in March 2013~\cite{grigorik-http-client-hints-00}. It became an IETF draft in November 2015~\cite{ietf-httpbis-client-hints-00} and then developed to an experimental RFC in February 2021~\cite{rfc8942}, but it is not an RFC standard yet. The specification of HTTP CHs was defined in a W3C draft community group report~\cite{weiss_client_2023}, but is not a W3C standard yet. Nevertheless, popular web browsers like Chrome and Edge already support HTTP CHs, which affects more than 75\% of web users worldwide~\cite{caniuse_client_2024}.

\subsection{Implementation in Web Browsers}\label{sec:background-implmentation}
HTTP CHs have become a fundamental feature in contemporary web browsers. Chrome first implemented support for HTTP CHs in version 85~\cite{chromium_user_2021}, which was released in August 2020~\cite{chrome_new_2020}. %
Microsoft Edge, built on the Chromium engine, also integrated support for HTTP CHs in alignment with Chrome. %
Similarly, Brave, also Chromium-based, synchronized its implementation with Chrome and Edge.
Safari and Firefox stand alone among the major browsers in not offering support for HTTP CHs. %

From this, it becomes apparent that HTTP CHs are supported by major browsers. The fact that this support is offered by the most widely used browsers with a desktop market share of 78\% (see Table~\ref{tab:client-hint-browser-support} and Section~\ref{sec:browser-support}) suggests that widespread availability can be assumed in practice. However, this development remains opaque to users because none of the supporting browsers offer users the ability to control HTTP CHs. For example, there is no way to limit third-party requests to low-entropy hints or to disable them altogether. In principle, this poses a significant potential for abuse, as the HTTP CHs that are originally intended to enhance privacy can inadvertently and possibly even unlawfully be exploited to track web users. Therefore, it is important to understand how HTTP CHs are used in practice in order to take appropriate measures to enhance transparency and control for web users.

\begin{table*}[t]
\centering
    \caption{Overview of HTTP CH support across different devices and Browsers.}
\begin{tabular}{@{}l | lll | lll | lll | lll | l | ll @{}}
\toprule
                             & \multicolumn{3}{|c|}{}                 & \multicolumn{3}{|c|}{}                 & \multicolumn{3}{|c|}{}               & \multicolumn{3}{|c|}{}       & Samsung      &  \multicolumn{2}{|c}{}                                 \\
Browser                             & \multicolumn{3}{|c|}{Chrome}                 & \multicolumn{3}{|c|}{Brave}                 & \multicolumn{3}{|c|}{Firefox}               & \multicolumn{3}{|c|}{Edge}       &  Internet      &  \multicolumn{2}{|c}{Safari}                                 \\
Platform                            & \macos                 & \ios                   & \android               & \macos                 & \ios                   & \android               & \macos                 & \ios                   & \android               & \macos                 & \ios                   & \android               & \android               & \macos                 & \ios                   \\
\midrule
User-Agent (Low Entropy)                           & \cmark & \xmark & \cmark & \cmark & \xmark & \xmark & \xmark & \xmark & \xmark & \cmark & \xmark & \cmark & \xmark & \xmark & \xmark \\
OS (Low Entropy)                  & \cmark & \xmark & \cmark & \cmark & \xmark & \xmark & \xmark & \xmark & \xmark & \cmark & \xmark & \cmark & \xmark & \xmark & \xmark \\
Prefers Mobile UX                    & \cmark & \xmark & \cmark & \cmark & \xmark & \xmark & \xmark & \xmark & \xmark & \cmark & \xmark & \cmark & \xmark & \xmark & \xmark \\
User Agent (High Entropy)              & \cmark & \xmark & \cmark & \xmark & \xmark & \xmark & \xmark & \xmark & \xmark & \cmark & \xmark & \cmark & \xmark & \xmark & \xmark \\
User Agent Brand List         & \cmark & \xmark & \cmark & \xmark & \xmark & \xmark & \xmark & \xmark & \xmark & \cmark & \xmark & \cmark & \xmark & \xmark & \xmark \\
OS (High Entropy)          & \cmark & \xmark & \cmark & \cmark & \xmark & \xmark & \xmark & \xmark & \xmark & \cmark & \xmark & \cmark & \xmark & \xmark & \xmark \\
Platform Architecture                      & \cmark & \xmark & \xmark & \xmark & \xmark & \xmark & \xmark & \xmark & \xmark & \cmark & \xmark & \cmark & \xmark & \xmark & \xmark \\
CPU Bitness                   & \cmark & \xmark & \xmark & \xmark & \xmark & \xmark & \xmark & \xmark & \xmark & \cmark & \xmark & \cmark & \xmark & \xmark & \xmark \\
Device Model                     & \cmark & \xmark & \cmark & \xmark & \xmark & \xmark & \xmark & \xmark & \xmark & \cmark & \xmark & \cmark & \xmark & \xmark & \xmark \\
Device Form Factor               & \xmark & \xmark & \xmark & \xmark & \xmark & \xmark & \xmark & \xmark & \xmark & \xmark & \xmark & \xmark & \xmark & \xmark & \xmark \\
Prefers Reduced Data                           & \xmark & \xmark & \xmark & \xmark & \xmark & \xmark & \xmark & \xmark & \xmark & \xmark & \xmark & \xmark & \xmark & \xmark & \xmark \\
Viewport Width                      & \cmark & \xmark & \cmark & \xmark & \xmark & \xmark & \xmark & \xmark & \xmark & \cmark & \xmark & \cmark & \cmark & \xmark & \xmark \\
Client DPR                         & \cmark & \xmark & \cmark & \xmark & \xmark & \xmark & \xmark & \xmark & \xmark & \cmark & \xmark & \cmark & \cmark & \xmark & \xmark \\
Client's RAM                       & \cmark & \xmark & \cmark & \xmark & \xmark & \xmark & \xmark & \xmark & \xmark & \cmark & \xmark & \cmark & \cmark & \xmark & \xmark \\
Round-Trip Time                                 & \cmark & \xmark & \cmark & \xmark & \xmark & \xmark & \xmark & \xmark & \xmark & \cmark & \xmark & \cmark & \cmark & \xmark & \xmark \\
Bandwidth                            & \cmark & \xmark & \cmark & \xmark & \xmark & \xmark & \xmark & \xmark & \xmark & \cmark & \xmark & \cmark & \cmark & \xmark & \xmark \\
Network Profile                                 & \cmark & \xmark & \cmark & \xmark & \xmark & \xmark & \xmark & \xmark & \xmark & \cmark & \xmark & \cmark & \cmark & \xmark & \xmark \\
Light/Dark Mode         & \cmark & \xmark & \xmark & \xmark & \xmark & \xmark & \xmark & \xmark & \xmark & \cmark & \xmark & \cmark & \cmark & \xmark & \xmark \\
Prefers Reduced Motion       & \xmark & \xmark & \xmark & \xmark & \xmark & \xmark & \xmark & \xmark & \xmark & \cmark & \xmark & \cmark & \cmark & \xmark & \xmark \\
Reduced Transparency & \xmark & \xmark & \xmark & \xmark & \xmark & \xmark & \xmark & \xmark & \xmark & \xmark & \xmark & \xmark & \xmark & \xmark & \xmark \\
Contrast Preference             & \xmark & \xmark & \xmark & \xmark & \xmark & \xmark & \xmark & \xmark & \xmark & \xmark & \xmark & \xmark & \xmark & \xmark & \xmark \\
Forced Colors                & \xmark & \xmark & \xmark & \xmark & \xmark & \xmark & \xmark & \xmark & \xmark & \xmark & \xmark & \xmark & \xmark & \xmark & \xmark \\
\bottomrule
\multicolumn{16}{@{}l@{}}{\macos: Desktop PC (macOS Catalina), \ios: iOS 16 (Mobile), \android: Android 10 (Mobile)}\\
\multicolumn{16}{@{}l@{}}{Versions: Chrome 116, Brave 116, Firefox 117 (iOS, Android) and 116 (macOS), Edge 116, Samsung Internet 111,}\\
\multicolumn{16}{@{}l@{}}{\hspace{3.7em} Safari 605 (macOS) and 604 (iOS)}\\
\end{tabular}
\label{tab:client-hint-browser-support}
\end{table*}

\section{Studying HTTP CH Usage}%
\label{sec:ch-in-the-wild}

To investigate HTTP CHs in the wild, we used various data collection and analysis methods. We outline them in the following, including ethical and legal considerations regarding the work.

\subsection{Extracting Historical Crawling Data}
The introduction of HTTP CHs goes back to the year 2013, which was long before we decided to study this measure. To trace the history of HTTP CHs in the web, we used data from the HTTP Archive~\cite{httparchive_faq_2023}. The data contains monthly to half-monthly crawls of millions of popular start pages on the Internet as determined by the Chrome User Experience Report (CrUX) data set~\cite{chrome_crux_2023}\footnote{The HTTP Archive integrated the CrUX URLs in January 2018. We did not find information on the URL data set used before that.}. The data used consisted of the URL, the timestamp of when it was crawled, and the received HTTP response headers for this crawl. A Chrome desktop browser with an empty browser cache crawled each URL from Google cloud instances inside the US~\cite{httparchive_faq_2023}. The crawler then stored the results in the data set. As the data sets for each crawl are quite huge (approx. 1~TB per crawl), we queried the data using Google's BigQuery~\cite{google_bigquery_2024}. In so doing, we extracted the first sent HTTP responses from all crawled websites that ever sent the \texttt{Accept-CH} HTTP response header. We focused on the first response to make sure that no HTTP CH headers were cached by the client at that time.

We started our data analysis with the December 2023 set and went back in time until we found no more HTTP CH headers in the data. As a result, we obtained the historical data of all start pages crawled by the HTTP Archive that ever used HTTP CHs.

\subsection{Data Collection on Login Pages}
The crawled historical data only included the start pages of websites. However, some websites may request other HTTP CH data during the login process, e.g. to prevent account takeover. Also, different privacy jurisdictions like GDPR~\cite{european_union_gdpr_2016} and CCPA~\cite{california_ccpa_2018} might limit the amount of data collected by online services, including HTTP CHs. For this reason, we additionally crawled HTTP CH header data from login pages of the Tranco list websites from four different regions on three different continents (North America: Johnstown, Ohio, USA; Europe: Frankfurt and Biere, Germany; Asia: Singapore) and two different ISPs, which were Amazon Web Services (AWS) and Deutsche Telekom (DT). We used this data to determine differences in the HTTP CH behavior compared to the start page, and across different regions and ISPs.

To determine login pages, we accessed the URLs from the Tranco 8M list from June 21st, 2022 and determined login page URLs using an automated process (see Section~\ref{sec:identifying-login-pages}). With the final list of login page URLs, we started the crawling process (see Section~\ref{sec:crawling-process}).

For the whole login page detection and crawling process, we used the Chromium browser version 103, which is compatible with HTTP CHs~\cite{caniuse_client_2024,chromium_user_2023}. As websites might use bot detection mechanisms~\cite{jonker_fingerprint_2019,wiefling_even_2019}, we automated the browser behavior with a patched version of the Puppeteer framework~\cite{wiefling_even_2019} to appear as a human-like user.

\subsubsection{Identifying Login Pages}
\label{sec:identifying-login-pages}
It is possible that websites request different client information in the login context. For instance, websites might not need the full browser version to generate basic traffic statistics. In case of RBA, e.g., a website might still be interested in the full browser version, as this might help to identify the legitimate user accessing an online account. Therefore, we decided to analyze both the start page and the login page of a website, to be able to spot differences.

The website start page appears when the URL of a Tranco URL entry is accessed. The login page is the web page where the login credentials (most commonly username and password) are requested by the website. This login page can be located at the same or a different URL as the start page, whereby the latter URL is not included in the Tranco list. Manually inspecting all 8M sites to determine the login page URL would be rather impractical, and also dangerous as some of them might contain illegal or harmful content. Therefore, based on our observations on popular websites, we created a systematic approach to determine the login URLs of the websites.

For each domain of the Tranco list, we conducted a three step process, which we describe in the following. We explain the process for the domain \texttt{example.com}. We considered Transport Layer Security (TLS) mandatory for login pages, as it protects login credentials from interception. Therefore, we only sent HTTP requests to the \texttt{https://} prefix version of a website. We ran this process on the server infrastructure of our university using a university IP address (location: Sankt Augustin, Germany).

\inlineheading{Step 1: Collecting Potential Login URLs}
As a first step, we determined a list of URLs that could potentially be a login URL. We designed this procedure based on our previous experiences with websites and content crawling from them. We tested and improved this procedure for one month, to make it as accurate as possible. We first collected three types of potential login URLs and chose the best candidate afterwards. These types were the following:

\begin{enumerate*}
    \item We opened \texttt{https://example.com}, waited until the page was fully loaded, and parsed the returned HyperText Markup Language (HTML)~\cite{berners-lee_hypertext_1995} code to allow further analysis. We then added all URLs contained in HTML \texttt{<a>} tags together with the link text to our candidate list. We assume that in most cases the login URL should be among these links, when a website has a login page.
    \item We sent an HTTP request to \texttt{https://example.com/login}, as we observed this pattern often at login pages of websites. When we received the HTTP status code \texttt{200 OK}, which indicates that the URL points to an existing resource, we added the URL with the page title retrieved from the HTML \texttt{<title>} tag---as we do not have a link text here---to our candidate list. This should have further increased the probability that we found a login page.
    \item As a last resort, we opened the Google search engine and entered the search query \texttt{login site:example.com}. This query should only return pages containing the word \texttt{login} at the \texttt{example.com} website. We added the results with the link text to our candidate list as well.
\end{enumerate*}

\inlineheading{Step 2: Scoring Login URLs}
After Step 1, we had a candidate list of potential login URLs. To identify the best candidate, we scored the URLs by checking both URL and the text associated with it. We defined the criteria after a discussion of two researchers who made observations at various websites in the wild.

We had a list of positive indicators, which consisted of different variations of the words \textit{``login''} and \textit{``sign in''} in German and English. We checked both languages, as we browsed from a German IP address, so language variations were possible. Based on the matches with those indicators, we gave different scores: Matches of both text and URL received three points, matching the URL but not the text received two points, and matching the text but not the URL received one point. We found the URL a better indicator than the link text for rare cases where websites displayed different languages than our tested ones.

We also found negative indicators. These were when the words \textit{``help''}, \textit{``premium''}, \textit{``pro''}, or \textit{``forgot password''} appeared in a link text or URL, with the URL being a higher indicator. We experienced that such URLs rather opened a registration, password recovery, or help desk page than a login page. We assumed that a higher number of negative indicators decreased the probability of a crawled URL being a login URL. Therefore, based on the matches with these negative indicators, we reduced the following values from the current scores using a ranking-based approach: Matches of both text and URL lost three points, matching URL but not the text lost two points, and matching text but not the URL lost one point.

\inlineheading{Step 3: Determining Best Candidate}
After the scoring process, we determined the best candidate for the login URL as the URL with the highest score. In case of multiple candidates with the highest score, we only focused on the one that appeared first in the list. We assume that links to login pages appear on the top area of a website, so this link would be the first in our list based on how we ordered it with our process.

With this approach, we determined 419K login page candidates on June 22nd, 2022, which we used for the crawling process. We verified the validity of our approach by reading the most popular URLs of the determined login pages, comparing them with our previous experiences on these websites, and checking a small subset of them whether they lead to the login page.

\subsubsection{Crawling Process}
\label{sec:crawling-process}
We crawled the login page candidates on a monthly basis, from August 7th, 2022 to December 21st, 2023\footnote{For technical reasons, we had crawling gaps of one (October 2022) and two months (October/November 2023). However, the impact should be minimal (see Section~\ref{sec:limitations}).}. We assumed that a monthly time interval should be sufficient to identify changes in the adoption of HTTP CHs. Also, as crawling a huge number of URLs produces a large amount of data, we had to limit our data collection to a reasonable level. 

For each login page URL, our crawler started a new Chromium browser session with empty cache and storage. We did this to make sure that no HTTP CH requests were cached. Then, Chromium initiated an HTTP request to the URL as a human-like Internet user, and the crawler recorded the \texttt{Accept-CH} value, if present, of the server's HTTP response. We also recorded all browser-initiated third party requests during the page loading, and recorded a present \texttt{Accept-CH} response as well.

To mitigate being detected as a bot, we crawled the data using a cluster of six different servers with six different IP addresses located in three different zones in the same data center. We furthermore limited the network throughput to 5 MBit/s and crawled the URLs in random order. Each server had 12 GB SSD, 4 GB RAM and one CPU core. Crawling the URLs took around three to four days per iteration.

\subsubsection{Filtering Login Pages}\label{sec:filtering-login-pages}

We aimed to investigate which websites request HTTP CHs on their own login page. Websites might change over time. To ensure a high data quality, we applied additional filtering after the crawling. Therefore, we made sure that the domain name of the login URL matched the domain name of the Tranco list entry. In so doing, e.g., \texttt{https://login.example.com} was considered the valid login URL of \texttt{https://example.com} while \texttt{https://login.example.org} was not.

\subsubsection{Extracting Third Party Domains}

We extracted the domain names from all crawled URLs and all their resources that were requested. To identify third party domains, we extracted the domain names from all URLs and then kept all URLs which did not belong to the same domain. We did this to only include URLs that are external resources that are likely not directly related to the original website (e.g., a subdomain pointing to an internal content delivery network).

\subsection{Data Analysis}

We used the following methodology to analyze our collected data.

\subsubsection{Aligning the Two Data Sets for Comparison}

To allow comparisons between the historical crawling and login pages crawling data sets, we only considered the domains in the analysis that were present in both of them. %
We took the data from December 2023 from both HTTP Archive and our login pages crawl, to ensure that we have the same state of crawling data. The crawling from the HTTP Archive was from December 13th to December 23rd, 2023 and the login pages crawling from December 16th to December 21st, 2023. As the total amount of HTTP CH websites did not change to a large degree in that month, we assume that the changes of websites using HTTP CHs were minimal.
We then extracted a union set of websites based on the domain names which were present in both data sets. We took the domain names as some websites have a dedicated subdomain for logins (e.g., \texttt{accounts.google.com} for \texttt{google.com}). Then, we compared the requested HTTP CHs for the start page and the login page. Our union subset contained 1,938 websites that used HTTP CHs.

\subsubsection{Identifying HTTP CHs}

We processed each \texttt{Accept-CH} HTTP response header sent by the websites. This header included the list of HTTP CHs requested by the server. The RFC~\cite{grigorik_http_2021} and W3C draft~\cite{weiss_client_2023} classified HTTP CHs into \textit{valid}, \textit{deprecated}, and \textit{experimental} ones (see Figure~\ref{fig:client-hint-occurences}). We took their classification and considered HTTP CHs \textit{not valid} when they did not appear in these documents. HTTP CHs classified as \textit{experimental} also belonged to the \textit{valid} ones.

\subsubsection{Determining Level of Detail}
To analyze the user trackability on websites, we rated each of the available valid and deprecated HTTP CHs by their level of detail that they potentially revealed about a user. Based on related work by Alaca and van Oorschot~\cite{alaca_device_2016}, and Wiefling et al.~\cite{Wiefling_Whats_2021}, we classified the HTTP CHs on a scale with the ratings \textit{very low}, \textit{low}, \textit{medium}, \textit{high}, and \textit{very high} (see Table~\ref{tab:client-hint-classification}). In cases in which both publications did not give a rating on a corresponding feature, we made the decision based on related features with similar expected entropy. Based on the classifications of all requested HTTP CHs, we determined the maximum level of distinguishing information per website.

{\newcommand{\W}{\cite{Wiefling_Whats_2021}}
\newcommand{\A}{\cite{alaca_device_2016}}
\newcommand{\WA}{\cite{Wiefling_Whats_2021,alaca_device_2016}}

\begin{table}[H]
  \centering
  \caption{Classification of the level of detail for each available HTTP CH (as defined in the RFC~\cite{grigorik_http_2021} and specification~\cite{weiss_client_2023})}
    \resizebox{\linewidth}{!}{
        \begin{tabular}{@{}lp{0.35\linewidth}llp{0.6\linewidth}@{}}
            \toprule
            \textbf{} & \textbf{} & \textbf{Level of} & \textbf{} & \textbf{} \\
            \textbf{HTTP CH} & \textbf{Header} & \textbf{Detail} & \textbf{Source} & \textbf{Explanation} \\
            \midrule
            \textbf{User Agent} &       &       &       &  \\
            \midrule
            User Agent (High Entropy) & Sec-CH-UA-Full-Version & Very High & \W     & User agent's full semantic version string \\
            User Agent Brand List & Sec-CH-UA-Full-Version-List & Very High & \W     & Full version for each brand in the user agent's brand list \\
            OS (High Entropy) & Sec-CH-UA-Platform-Version & High & \WA     & Operating system version \\
            Device Model & Sec-CH-UA-Model & Medium & \A & Device model \\
            OS (Low Entropy) & Sec-CH-UA-Platform & Low   & -     & Operating system (low entropy hint) \\
            User Agent (Low Entropy) & Sec-CH-UA & Low   & \W     & Branding and major version \\
            CPU Bitness & Sec-CH-UA-Bitness & Very Low & -     & CPU architecture bitness (mostly 32 or 64 bit) \\
            Device Form Factor & Sec-CH-UA-Form-Factor & Very Low & \W     & Form factor of device (\texttt{Automotive}, \texttt{Mobile}, \texttt{Tablet}, \texttt{TV}, \texttt{VR}, \texttt{XR}, \texttt{Unknown}) \\
            Is Windows64 & Sec-CH-UA-WoW64 & Very Low & -     & Binary runs on 64-bit Windows \\
            Platform Architecture & Sec-CH-UA-Arch & Very Low & -     & Platform architecture (mostly x86 or ARM) \\
            Prefers Mobile UX & Sec-CH-UA-Mobile & Very Low & -     & Prefers a mobile user experience (boolean value) \\
            \midrule
            \textbf{User Preference Media} &       &       &       &  \\
            \midrule
            Contrast Preference & Sec-CH-Prefers-Contrast & Very Low & -     & Preference for contrast \\
            Forced Colors & Sec-CH-Forced-Colors & Very Low & -     & Forces a color scheme \\
            Light/Dark Mode & Sec-CH-Prefers-Color-Scheme & Very Low & -     & Prefers light or dark color scheme \\
            Prefers Reduced Motion & Sec-CH-Prefers-Reduced-Motion & Very Low & -     & Reduced motion preference setting (either \texttt{no-preference} or \texttt{reduced}) \\
            Reduced Transparency & Sec-CH-Prefers-Reduced-Transparency & Very Low & -     & Prefers reduced transparency \\
            \midrule
            \textbf{Device Information} &       &       &       &  \\
            \midrule
            Viewport Width & Viewport-Width & Very High & \W     & Layout viewport width \\
            Width & Width & Very High & \W     & Desired resource width \\
            Client DPR & Content-DPR & Low   & \W     & Image device pixel ratio \\
            Client's RAM & Device-Memory & Low   & \WA   & Approximate amount of available client RAM memory \\
            Image DPR & DPR   & Low   & \W     & Client device pixel ratio \\
            \midrule
            \textbf{Network} &       &       &       &  \\
            \midrule
            Bandwidth & Downlink & High  & \A     & Approximate bandwidth of the client's connection to the server \\
            Network Profile & ECT   & High  & \A     & The effective connection type ("network profile") that best matches the connection's latency and bandwidth \\
            Round-Trip Time & RTT   & High  & \W     & Application layer round trip time in milliseconds \\
            Prefers Reduced Data & Save-Data & Very Low & -     & Preference for reduced data usage (boolean value) \\
            \bottomrule
        \end{tabular}%
    }
  \label{tab:client-hint-classification}%
\end{table}
}

\subsubsection{Website Classification}

We classified the different HTTP CH websites into different categories. %
Inspired by related work~\cite{gavazzi_a_2023}, we queried the McAfee URL classification API~\cite{mcafee_check_2023} for all the websites that used HTTP CHs.

Gavazzi et al.~\cite{gavazzi_a_2023} studied more than 200 popular websites inside the Tranco 5K regarding their RBA usage. We obtained their data set with permission, and used their results to classify the crawled websites from the Tranco 5K into those that used some form of RBA and those that likely did not use RBA.

From the third party requests, we identified known trackers using the EasyPrivacy list~\cite{easylist_easylist_2024} that is commonly integrated in adblockers.

\subsubsection{Comparisons}

To identify significant differences between the HTTP CH usage of different website categories and types in RQ2, we used Pearson's Chi-squared test for contingency table analysis ($\chi^2$). We considered $p < 0.05$ as significant. For pairwise comparisons, we adjusted the p-values using Bonferroni correction and used $\alpha = 0.05$ as the start level.

\subsection{Ethical and Legal Considerations}

We do not have a formal institutional review board process at our university. Nevertheless, this study was carefully reviewed and approved by the university's data protection officer. Furthermore, we made sure to minimize potential harm by complying with the ethics code of the German Sociological Association (DGS) and the standards of good scientific practice of the German Research Foundation (DFG). We also made sure to comply with the GDPR~\cite{european_union_gdpr_2016}. 

We used an external API to classify websites. Since the number of URLs was much lower than the original 8M URLs used for crawling, we kept the traffic to this external API to a minimum. Furthermore, we only sent one HTTP request to each website in the crawling process to keep the traffic as low as possible. We did not record any response data beyond the \texttt{Accept-CH} HTTP header, to make sure that no personally identifiable information was in the collected data. We also identified the login page URLs with an automated process to avoid that a person was accidentally confronted with harmful or illegal content.

\subsection{Results}

\begin{figure}[t]
    \centering
    \includegraphics[width=0.9\linewidth]{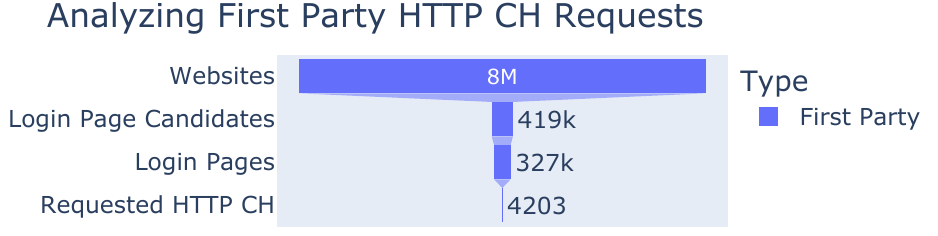}\\
    \includegraphics[width=0.9\linewidth]{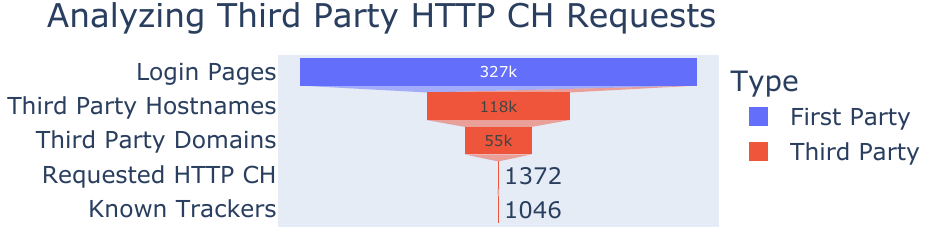}
    \caption{Overview of the collected data and the results of our data analysis, based on the data from December 2023.}
    \label{fig:analysis-filtering}
\end{figure}

In the following sections, we show the results of our study ordered by the research questions. Unless otherwise noted, we used the results from the European region using the DT ISP. Figure~\ref{fig:analysis-filtering} shows the amount of data we collected and the identified websites and third parties that requested HTTP CHs during our study.

\section{Adoption of HTTP CHs (RQ1)}

We first describe the results regarding the HTTP CH adoption on website start pages and login pages, also those using RBA.

\subsection{Website Start Pages (RQ1a)}

\begin{figure}[t]
    \centering
    \includegraphics[width=\linewidth]{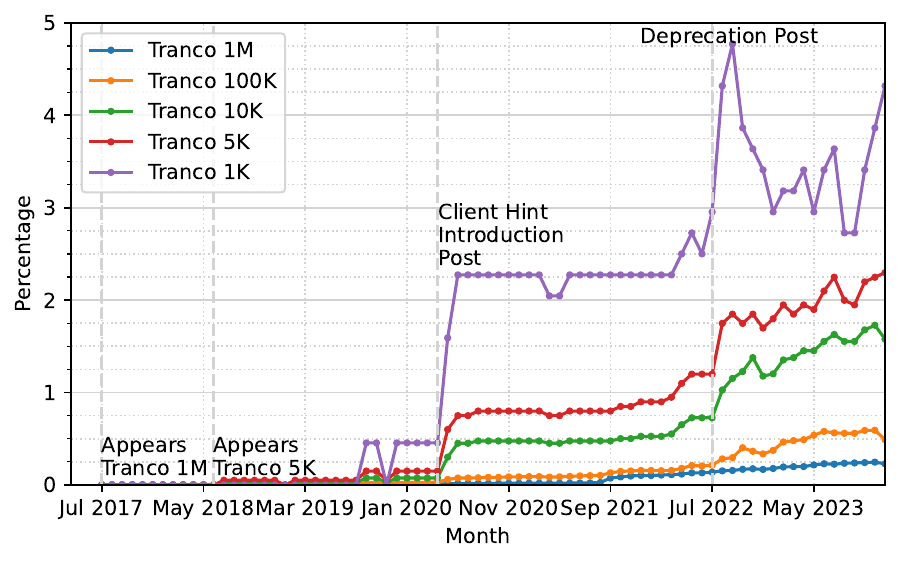}
    \caption{Observed HTTP CH adoption rates on start pages over time, grouped by rankings inside the Tranco list. The data is taken from the HTTP Archive crawling data, that crawled the whole Web each month.%
    }
    \label{fig:client-hint-usage-httparchive-data}
\end{figure}

The adoption of HTTP CHs on website start pages has increased since its introduction in 2013, with some notable peaks in 2020 and 2022 (see Figure~\ref{fig:client-hint-usage-httparchive-data}). An HTTP CHs request first appeared on the \texttt{punknews.org} website in July 2017, and this was the only website using HTTP CHs for a long time. From July 2018, more websites started adopting HTTP CHs, and this was also the first time they appeared inside the Tranco 5K. The Tranco 1K followed a year after in September 2019.

In June 2020, Google wrote a blog article about HTTP CHs with the aim of soon replacing UAS with HTTP CHs in the Chrome browser~\cite{merewood_improving_2020}. After that, a first large peak in HTTP CH adoption was noticable. At the end of August 2022, the Google Chrome team also warned developers that the UAS would change to a reduced version in October 2022 and that websites would have to switch to HTTP CHs to gather more user information again~\cite{ney_prepare_2022}. At this point, we observed the second big increase in HTTP CH adoption on the Web. Overall, the use of HTTP CHs has remained at a very marginal level since its introduction.

\subsection{Difference Between Login Page and Start Page (RQ1b)}
For the crawled websites that use HTTP CHs, 54\% behaved differently on start and login pages. The majority of them (90.6\%) did not request HTTP CHs on their login page while requesting HTTP CHs on the start page. However, the login pages requested more HTTP CHs ($\widetilde{X}=4.5$; $SD=2.86$) than the start pages ($\widetilde{X}=1$; $SD=3.43$).

\begin{figure}[t]
    \centering
    \includegraphics[width=0.495\linewidth]{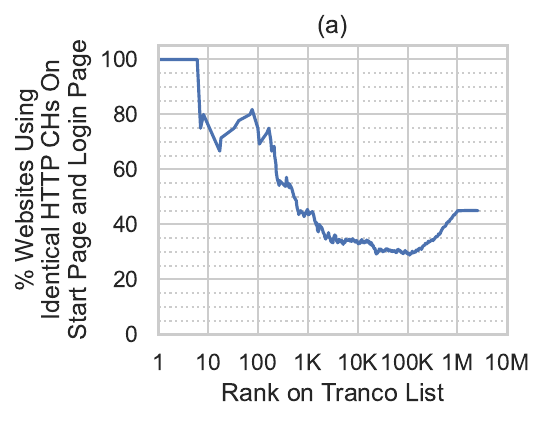}
    \includegraphics[width=0.495\linewidth]{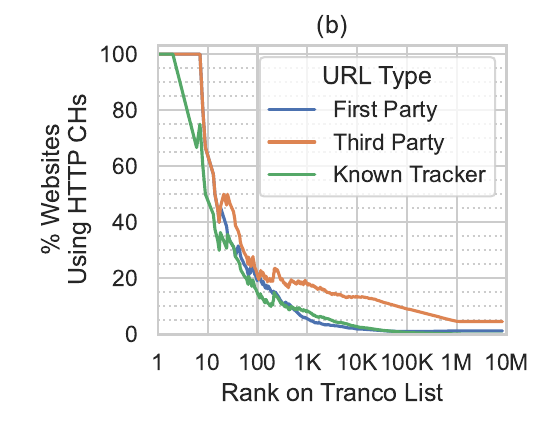}
    \caption{(a) Percentage of websites showing identical HTTP CH behavior on their login page compared to the start page. (b) Percentage of websites that use HTTP CHs based on their rankings inside the Tranco list. We calculated the usage based on the data from December 2023.}
    \label{fig:tranco_rank_to_identical_ch_usage}
        \label{fig:tranco-rank-to-ch-usage}
\end{figure}

The behavior was not constant based on the Tranco ranking (see Figure~\ref{fig:tranco_rank_to_identical_ch_usage}a). While the majority of the 10 most popular websites did not show a different behavior, few websites until Tranco 100 requested different HTTP CHs. Websites between Tranco 100 and Tranco 1K then often requested different HTTP CHs. After that, we observed mixed behavior until between Tranco 10K and Tranco 100K. After that, most websites requested the same HTTP CHs on login page and start page again.

As the HTTP CHs requested by login pages seemed to be more detailed, we used our login pages data for the subsequent analyses. This allowed us to gain a larger and probably more realistic view of the HTTP CH adoption on the Web.

\subsection{Login Pages Using RBA (RQ1c)}

The amount of RBA websites that used HTTP CHs during our observation was low. Out of 73 websites with potential RBA usage, 12 (16.4\%) sent an HTTP CH header. In terms of different online service providers, these were nine, while there were multiple domains belonging to the online service Amazon%
. We show more details on these services and their requested HTTP CHs in Section~\ref{sec:rba-websites}.

\subsection{Distribution in Practice (RQ1d)}

Aside from very popular online services, the overall HTTP CH adoption on login pages on the Web remained very low. While the the most popular websites from the Tranco list started using HTTP CHs, the adoption rate constantly declined with a lower rank on the Tranco list (see Figure~\ref{fig:tranco-rank-to-ch-usage}b). However, many of the websites included requests to third party domains, some of which also requested HTTP CHs. When taking these into account, HTTP CH data was collected by third parties on 18.1\% of websites inside the Tranco~1K. About half of them were also known trackers at this point.

\subsection{Discussion}

Websites tended to requested more HTTP CHs on their login pages than on their start pages. It could be possible that they only requested additional information when it was necessary for their service (e.g., preventing attacks by malicious actors). However, it could also be possible that they hid their real HTTP CH usage from crawlers browsing the start pages only. Nevertheless, this suggests that the typical crawl of a website's start page may not reflect real-world HTTP CH usage and that the collected user information could be more detailed. Following this, future research should include more URLs besides a website's start page to obtain more realistic usage statistics on HTTP CHs.

HTTP CHs seemed not to be used by a lot of websites in practice. However, the websites that used HTTP CHs were very influential on people's daily lives and have a large amount of traffic (e.g., Google, Facebook, and Amazon). Therefore, they could potentially still collect a lot of user data although the HTTP CHs were introduced as a measure to protect \textit{``users' privacy [...] against covert tracking methods''}~\cite{merewood_improving_2020}. This is also highlighted by the large amount of embedded third party resources requesting HTTP CHs. We did not identify all of them as trackers. Nevertheless, we assume that this is only a lower bound as trackers try to circumvent detection~\cite{le_cv-inspector_2021}.

\section{Requested HTTP CHs (RQ2)}

Requested HTTP CHs can vary largely in practice. In the following, we show the differences between the various website types and client attributes, as well as the level of detail that websites requested.

\subsection{Used HTTP CHs (RQ2a, b)}

\begin{figure*}[t]
    \centering
    \includegraphics[width=\linewidth]{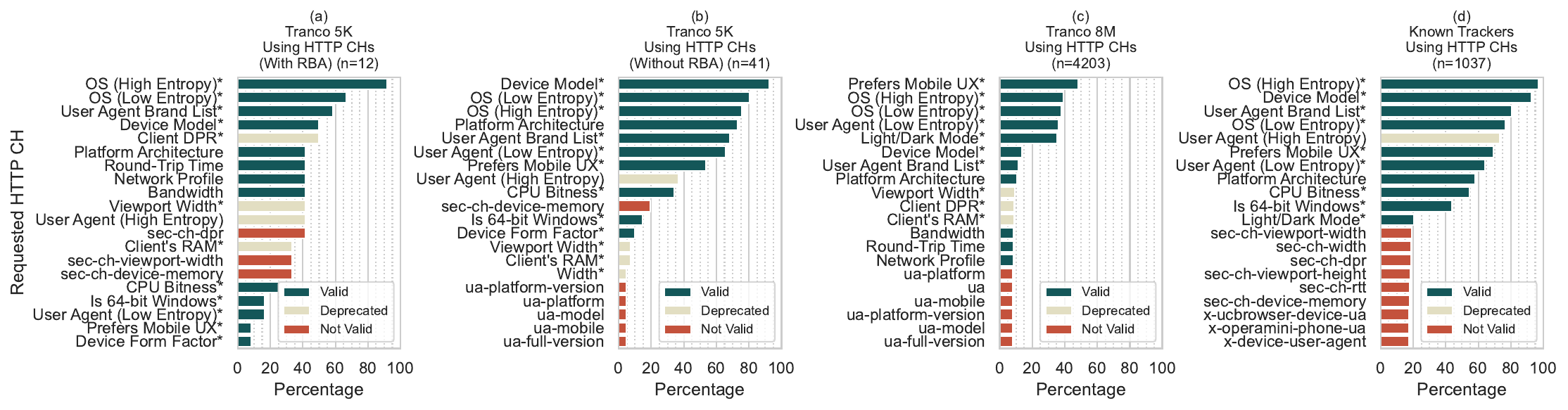}
    \caption{Top 20 HTTP CHs used among all websites and trackers that used HTTP CHs in the study. (*: Experimental HTTP CH)} 
    \label{fig:client-hint-occurences}
\end{figure*}

\begin{table}[t]
    \centering
    \caption{Significant $\chi ^2$ results and pairwise comparison $p$-values for valid HTTP CHs. We omitted $p$-values greater than 0.05.}
    \resizebox{\linewidth}{!}{
    \begin{tabular}{@{}ll | rl | rrr@{}}
    \toprule
               {} & Level of &  &  & 5K with RBA/  & Tranco 8M / & Tranco 8M /   \\  
    HTTP CH &  Detail & $p$ & $\chi ^2$ & 5K without RBA & 5K with RBA & 5K without RBA \\
    \midrule
    Viewport Width            &       Very High &  0.0006 &  14.8 &                                             0.0411 &                                             0.0027 &                                                  - \\
    User Agent Brand List     &       Very High &  $<$0.0001 & 146.9 &                                                  - &                                             $<$0.0001 &                                             $<$0.0001 \\
    User Agent (HE) &       Very High &  $<$0.0001 & 104.4 &                                                  - &                                             $<$0.0001 &                                             $<$0.0001 \\
    Bandwidth                 &            High &  0.0001 &  18.6 &                                             0.0034 &                                             0.0012 &                                                  - \\
    OS (HE)         &            High &  $<$0.0001 &  36.4 &                                                  - &                                             0.0018 &                                             $<$0.0001 \\
    Network Profile           &            High &  0.0001 &  18.7 &                                             0.0034 &                                             0.0011 &                                                  - \\
    Round-Trip Time           &            High &  0.0001 &  18.6 &                                             0.0034 &                                             0.0012 &                                                  - \\
    Device Model              &          Medium &  $<$0.0001 & 216.1 &                                             0.0074 &                                             0.0038 &                                             $<$0.0001 \\
    OS (LE)          &             Low &  $<$0.0001 &  35.1 &                                                  - &                                                  - &                                             $<$0.0001 \\
    Client's RAM              &             Low &  0.0113 &   9.0 &                                                  - &                                             0.0412 &                                                  - \\
    User Agent (LE)  &             Low &  0.0002 &  17.4 &                                             0.0220 &                                                  - &                                             0.0005 \\
    Client DPR                &             Low &  $<$0.0001 &  26.7 &                                             0.0004 &                                             $<$0.0001 &                                                  - \\
    Device Form Factor        &        Very Low &  $<$0.0001 & 110.6 &                                                  - &                                             0.0474 &                                             $<$0.0001 \\
    Platform Architecture     &        Very Low &  $<$0.0001 & 166.6 &                                                  - &                                             0.0090 &                                             $<$0.0001 \\
    Prefers Mobile UX         &        Very Low &  0.0175 &   8.1 &                                             0.0422 &                                             0.0417 &                                                  - \\
    Light/Dark Mode           &        Very Low &  $<$0.0001 &  22.8 &                                                  - &                                                  - &                                             0.0001 \\
    Prefers Reduced Data      &        Very Low &  0.0007 &  14.5 &                                                  - &                                                  - &                                                  - \\
    Is 64-bit Windows         &        Very Low &  $<$0.0001 &  57.2 &                                                  - &                                             0.0074 &                                             $<$0.0001 \\
    CPU Bitness               &        Very Low &  $<$0.0001 &  79.4 &                                                  - &                                             0.0336 &                                             $<$0.0001 \\
    \bottomrule
    \multicolumn{5}{@{}l}{HE: High Entropy, LE: Low Entropy}
    \end{tabular}
    }
    \label{tab:valid-client-hints-significance}
\end{table}

The usage of HTTP CHs on Tranco 5K websites significantly differed from the full Tranco 8M crawled (see Figure~\ref{fig:client-hint-occurences}b and c). There were also significant differences between Tranco 5K websites with and without RBA usage. Especially the \textit{high entropy user agent} feature was significantly less used on the Tranco 8M than on the Tranco 5K. Table~\ref{tab:valid-client-hints-significance} shows the HTTP CHs that were signficantly different among the website types, with the Tranco 5K collecting more high-entropy user data than the Tranco 8M. Some websites also requested invalid HTTP CHs, mostly because of typos or potential misinterpretation of the HTTP CH values (see Section~\ref{sec:requested-chs-discussion}).

Some HTTP CHs were significantly less popular among known trackers ($p<0.0001$, see Figure~\ref{fig:client-hint-occurences}c and d). These were \textit{(Viewport) Width}, \textit{Bandwidth}, \textit{Network Profile}, \textit{Round-Trip Time}, \textit{Image DPR}, \textit{Client DPR}, \textit{Prefers Reduced Data}, and \textit{Device Form Factor}. The other HTTP CHs were not significantly different from the Tranco 8M.

\subsection{Geographical Location and ISP (RQ2c)}

We observed differences between regions. While our crawlings from North America and Asia both identified 4,077 websites requesting HTTP CHs, this was different in our Europe crawls. The crawl using the AWS ISP revealed 4,089 websites and the crawl with the DT ISP identified 4,203 websites. The ranking of requested HTTP CHs did not change between the different crawls.

\subsection{Level of Detail Requested (RQ2d)}

\begin{figure}
    \centering
    \includegraphics[width=\linewidth]{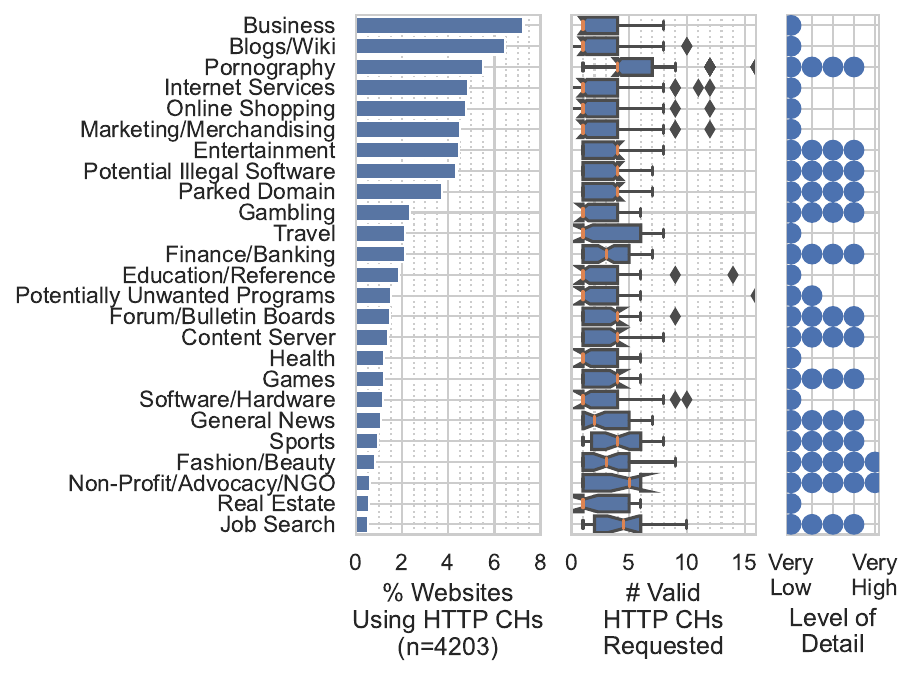}
    \caption{Top 25 categories of the Tranco 8M websites that used HTTP CHs, their percentage, their number of valid HTTP CHs requested, and the median level of detail they requested with the HTTP CHs. There is a maximum of 19 possible HTTP CHs. We also counted deprecated HTTP CHs as valid here, as they can still be valid in some browsers.}
    \label{fig:website-categories}
\end{figure}

Figure~\ref{fig:website-categories} shows the median requested level of detail for each website category. Many of the crawled website categories requested a high level of detail, like the fine-grained browser or OS name and version. These were also the ones that requested more than the median of one HTTP CH.

\subsection{RBA Websites (RQ2e)}\label{sec:rba-websites}

Figure~\ref{fig:client-hint-occurences}a shows the most requested HTTP CHs from the websites that used RBA. Compared to the Tranco 5K without RBA, they significantly requested more HTTP CHs with high level of detail (see Table~\ref{tab:valid-client-hints-significance}).

Table~\ref{tab:rba-websites-requested-client-hints} lists the identified RBA-using websites and the information they requested from the user with the HTTP CHs. Except for Amazon, all websites requested user agent information. Amazon and Etsy, however, requested network information like the bandwidth and the client-originated RTT. These were also the RBA-using websites that requested the highest level of detail.

\begin{table}[]
    \centering
        \caption{Overview of RBA websites that use HTTP CHs, and the HTTP CHs they requested (sorted from very high to very low level of detail).}
        \resizebox{\linewidth}{!}{
    \begin{tabular}{@{}p{0.25\linewidth} p{0.25\linewidth} p{0.4\linewidth} p{0.3\linewidth} p{0.4\linewidth}@{}}
        \toprule
        \textbf{Website} &  \textbf{Classification} &                                                         \textbf{(Very) High} &                           \textbf{Medium} &                                                         \textbf{(Very) Low} \\
        \midrule
        amazon.[com, in, de, co.uk]     & Online Shopping	    &  OS (High Entropy)*\newline Viewport Width\newline Round-Trip Time\newline Bandwidth\newline Network Profile &  & OS (Low Entropy)*\newline Client DPR\newline Client's RAM\\
        bedbathand\linebreak beyond.com & Online Shopping & OS (High Entropy)\newline User Agent Brand List & Device Model & Platform Architecture \\
        etsy.com & Online Shopping & OS (High Entropy)\newline User Agent Brand List\newline Round-Trip Time\newline Bandwidth\newline Network Profile &  & Prefers Mobile UX\newline User Agent (Low Entropy)\newline Platform Architecture\newline CPU Bitness\newline OS (Low Entropy)\newline Client DPR\newline Prefers Reduced Data \\
        facebook.com       & Social Networking	                                                & OS (High Entropy)\newline User Agent Brand List\newline Viewport Width & Device Model & Light/Dark Mode\newline Client DPR \\
        google.com              & Internet Services	                                      & OS (High Entropy)\newline User Agent Brand List\newline User Agent (High Entropy) & Device Model & Is 64-bit Windows\newline Platform Architecture\newline CPU Bitness\newline OS (Low Entropy)\newline Device Form Factor \\
        mail.ru & Portal Sites & OS (High Entropy)\newline User Agent Brand List\newline User Agent (High Entropy) & Device Model & OS (Low Entropy) \\
        microsoft.com              & Business &                                    OS (High Entropy),\newline User Agent (High Entropy) &       -       &                                                                             OS (Low Entropy) \\
        paypal.com                                            &   Finance/ Banking	&             OS (High Entropy)\newline User Agent Brand List\newline User Agent (High Entropy) & Device Model & Is 64-bit Windows\newline Platform Architecture\newline CPU Bitness \\
        p*****b.com                & Pornography                                        &  OS (High Entropy)\newline User Agent Brand List\newline User Agent (High Entropy) & Device Model & User Agent (Low Entropy)\newline Platform Architecture\newline OS (Low Entropy) \\
        \bottomrule
        \multicolumn{5}{@{}l}{*: Not on amazon.com}\\
    \end{tabular}
    }
    \label{tab:rba-websites-requested-client-hints}
\end{table}

\subsection{Discussion}
\label{sec:requested-chs-discussion}

Most websites tended to collect low entropy data, except for the OS where high and low entropy versions were requested at the same level. We assume that strong privacy regulations~\cite{european_union_gdpr_2016,california_ccpa_2018} might had an impact here, so that online services avoided collecting high entropy data without explicit user consent. This could be an improvement of privacy compared to the UAS. %
Nevertheless, Tranco 5K websites tended to collect more fine-grained user data than the Tranco 8M. That is probably because their business models often depend mainly on collected user data~\cite{zuboff_surveillance_2019} and that they have enough money and power to defend themselves against privacy lawsuits compared to smaller websites~\cite{wiefling_data_2022}. Also, websites might still be able to gather user information from third parties although they do not collect this information themselves, e.g, by matching it with the full IP address that can still be recorded.

In contrast to the other websites, Amazon did not seem to rely on the high-entropy user agent for their RBA algorithm. Instead, they collected network information like the RTT. A more reliable version would be to measure the RTT from the server side, which is possible, e.g., with WebSockets~\cite{Wiefling_Whats_2021} or token exchanges~\cite{han_detecting_2023}. Nevertheless, it could be possible that the server-side RTT is also measured, and then compared with the client-submitted version to detect whether the client tried to spoof some values%
.

Our crawlings revealed that some websites change their HTTP CH behavior based on ISP and region. Especially the different ISP revealed more than 100 HTTP CH using websites than before. This likely means that some websites detected AWS instances and changed their behavior based on it. Therefore, we assume that our results using the different ISP better reflect the real-world HTTP CH behavior on the Web. Also, we were able to detect more websites with HTTP CHs when residing in the EU. One reason for this could be that we identified the login pages from a server inside one country and that few websites applied some kind of geoblocking to clients outside of that country. Based on our crawled data, however, we cannot fully test this hypothesis as the access to the server-side logic would have been required.

HTTP CHs are not consistently worded. When related to the UAS, they start with \texttt{sec-ch-}. In all the other cases, for device and network information, they do not have this prepending string. We had the impression that some website developers seemed to get this wrong, even those belonging to high-traffic websites inside the Tranco 5K. This can be seen by the fact that the HTTP CH for client's RAM (\texttt{Device-Memory}) was often spelled wrong (\texttt{Sec-CH-Device-Memory}, see Figure~\ref{fig:client-hint-occurences}a, b, and d). Further research should investigate whether this is a common usability issue for developers and how HTTP CHs should be designed to avoid such pitfalls.

\section{Impact (RQ3)}

To better understand the impact of our results, we also studied the interconnection of the different third party domains and the browser support in practice.

\subsection{Interconnection of Third Party Domains (RQ3a)}

\begin{figure}
    \centering
    \includegraphics[width=\linewidth]{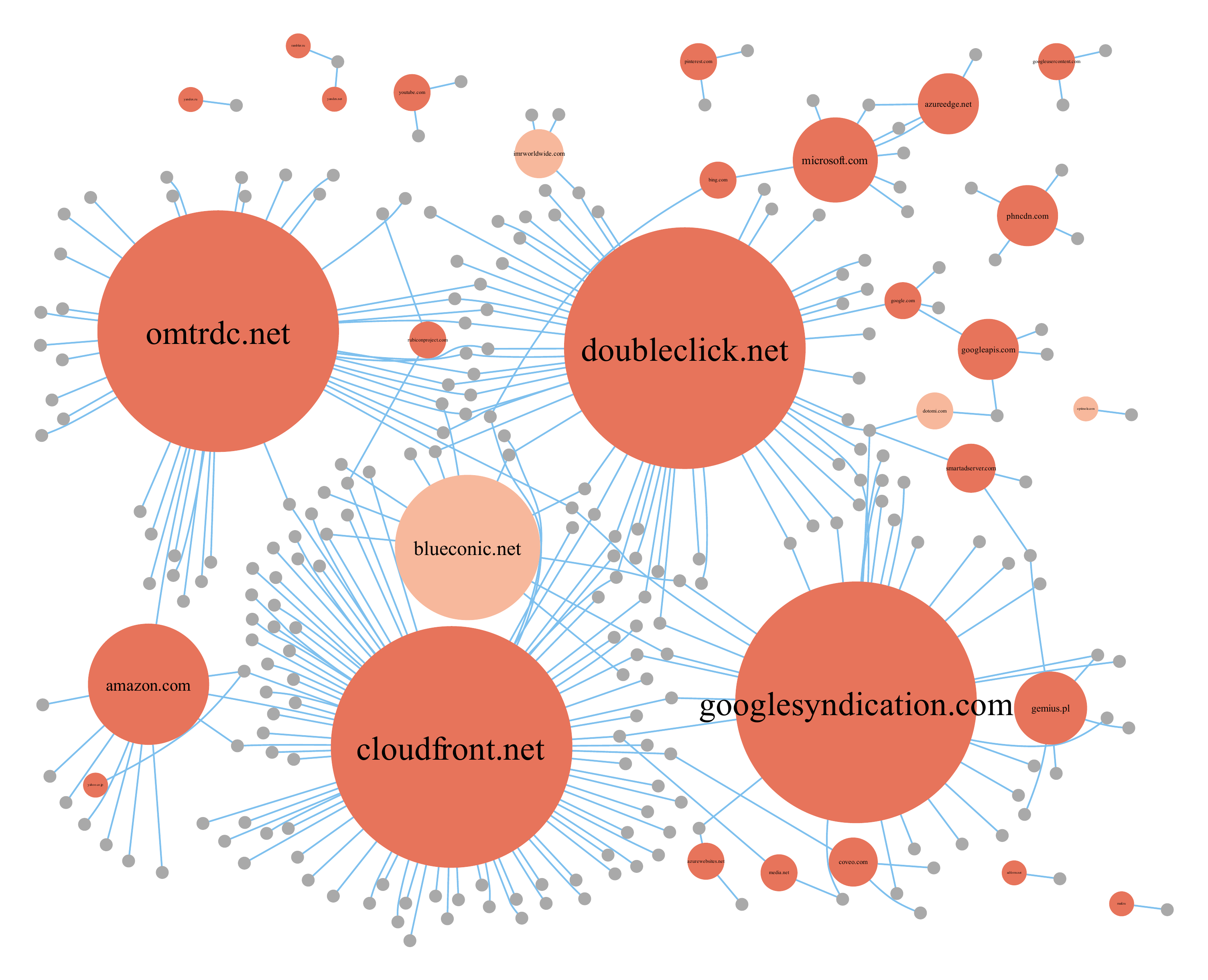}
    \caption{Network of third party domains that request HTTP CH data at the login pages of Tranco 5K websites (dark gray dots). The color at the third party domains shows their requested level of detail (light red: high, dark red: very high). None of the third party domains requested a medium or lower level of detail from the user}
    \label{fig:network-of-third-party-domains}
\end{figure}

Some crawled login pages implicitly submitted HTTP CH data to large tracking and content delivery networks by referencing third parties, who were integrated on other login pages as well (see Figure~\ref{fig:network-of-third-party-domains}). These third party domains were interconnected with popular websites, many of which did not use HTTP CHs themselves. These covered 18.1\% of the crawled Tranco 1K, 13.8\% of Tranco 5K, 13.3\% of Tranco 10K, and 9.1\% of Tranco 100K websites (see Figure~\ref{fig:tranco-rank-to-ch-usage}b).

\subsection{Browser Support (RQ3b)}
\label{sec:browser-support}

To estimate the real-world impact of HTTP CHs, we measured how browsers on three different devices (desktop, iOS, Android) interpret the different HTTP CHs. Therefore, we opened a website that sent an \texttt{Accept-CH} HTTP header containing all possible values from the RFC~\cite{grigorik_http_2021} and specification~\cite{weiss_client_2023}. Then we measured the response of the corresponding browser. We included the most popular browsers based on their market share in desktop and mobile devices~\cite{statscounter_market_2024}. As a privacy-baseline besides Firefox, we also included Brave, which is a well-known privacy-oriented browser~\cite{leith_web_2021}.
 
The results show that web browsers Chrome and Edge support a wide range of HTTP CHs in the desktop and Android browser versions (see Table~\ref{tab:client-hint-browser-support} in Section~\ref{sec:background-implmentation}). Brave on desktop and Samsung Internet for Android only support a subset of them. All mobile browsers on iOS as well as the Safari and Firefox browsers on all devices did not support any HTTP CH variants.

\subsection{Discussion}

Although the amount of third parties requesting HTTP CHs was low compared to the first parties, their influence on tracking users across websites is huge.

Most of the third parties requested a very high level of information using HTTP CHs, potentially allowing them to track users across websites. Some involved third parties were online services, which were also interconnected with trackers (e.g., amazon.com, microsoft.com, google.com). With a successful login on these online services, we assume that the involved third parties might be able to associate the tracking data with a user account and eventually also a real-world identity. This could be done with approaches like cookie respawning~\cite{fouad_my_2022}, which aim to restore cookie data with the help of browser fingerprinting attributes. This affects the majority of web users. Following the market shares of Chrome (65\% desktop, 65\% mobile), Edge (13\% desktop, 0.3\% mobile), and Samsung Internet (4\% mobile)~\cite{statscounter_market_2024}, we can assume that at least 78\% of desktop and 69\% of mobile users are trackable with HTTP CHs.

Interestingly, browsers on iOS do not seem to send HTTP CH responses. One reason for this could be that app developers have to meet high data protection requirements in order to get into Apple's App Store~\cite{apple_user_2024}.

\section{Related Work}
The invasion of web user privacy by third parties while browsing the Web is ubiquitous today. Over the years, many tracking technologies have evolved, with cookies being the first and still widely used tracking technology~\cite{bujilow_a_2017,sanchez-rola_can_2019,solomos_clash_2020,fouad_my_2022}. With the many countermeasures developed and implemented in web browsers, either as an integral part or as available add-ons~\cite{sanchez-rola_web_2017,schoni_block_2024}, new approaches to track web users evolved. These include user sessions~\cite{dao_cname_2021,acar_the_2014}, client memory and cache~\cite{solomos_tales_2021}, domain name system entries~\cite{dao_characterizing_2020,demir_towards_2022}, and fingerprinting~\cite{laperdrix_browser_2020}. Some of them are even rather persistent tracking mechanisms~\cite{acar_the_2014,fouad_my_2022}. The UAS has always been part of this tracking mixture. However, the UAS can also be used for good, for example in RBA where it is considered a useful feature to detect and prevent malicious account takeovers~\cite{Wiefling_Whats_2021,Wiefling_Pump_2022}. HTTP CHs were introduced to replace the UAS.%

Senol and Acar~\cite{senol_unveiling_2023} measured the usage of HTTP CHs for tracking purposes on the top 100K sites from the Chrome User Experience Report list in June 2023. They found that third-party domains frequently requested high entropy HTTP CHs from the user. Our study confirms their observations. Our results go beyond that, and provide a timely overview of HTTP CH usage on the Web, investigated the HTTP CHs usage on (RBA-instrumented) login pages, and identified differences in HTTP CH usage between first and third party domains, start pages and login pages, geographical user locations, and used ISPs.

Intumwayase et al.~\cite{intumwayase_ua-radar_2023} crawled a rank-sliced randomized sample of 12K domains belonging to the Tranco 1M to study the usage of UASs on the Web. They found that only few websites use the user agent to adjust the website contents. Following the results, they suggested to replace the UAS with HTTP CHs to reduce trackability by online services. Our analysis showed, however, that trackers seem to adjust to this change and obtain even more fine-grained user information via HTTP CHs as well.

Related work also measured the usage of tracking mechanisms on the web. 
Papadogiannakis et al.~\cite{papadogiannakis_user_2021} crawled one million websites to study whether they continue tracking users that rejected a cookie consent banner. They found that more than 14K of them tracked their users even when they did not consent to cookie-based tracking.
Fouad et al.~\cite{fouad_my_2022} crawled the top 30K websites and found that more than 1K of them use browser fingerprinting to restore cookies deleted by a user. Especially the IP address and user agent features were commonly used to achieve that.
In contrast to our work, all studies did not include HTTP CHs. Our work furthermore shows that HTTP CHs are commonly used by known trackers, and therefore potentially used to improve the user tracking abilities.

Iqbal et al.~\cite{iqbal_fingerprinting_2021} measured the usage of browser fingerprinting on the Alexa 100K websites. Bahrami et al.~\cite{bahrami_fp-radar_2022} did a similar approach with long-term measurement by crawling historical data from the Wayback Machine~\cite{wayback_machine_2023}. Both works proposed tooling to detect usage of different JavaScript API components for tracking.
However, as HTTP headers were out of scope, they did not mention HTTP CHs in their work. Also, the methodology using the Wayback Machine has flaws as it crawls from different locations and different time periods~\cite{hantke_you_2023}. Using our own crawling, we showed that HTTP CHs are probably used by trackers for tracking purposes.

\section{Limitations}\label{sec:limitations}
Our results are limited to the points in time where we crawled the data. It is possible that some websites requested HTTP CHs in a time period that was shorter than one month. Judging from the observed overall tendencies regarding HTTP CH adoption over time, however, we assume that the impact of this should be minimal.

We did not interact with the login pages and only recorded the first HTTP response. It is still possible that websites would have requested HTTP-CH data after submitting the login form. Nevertheless, our data still provides indications that websites show a different HTTP CH behavior on login pages compared to their start pages.

Our goal was to measure how web browsers with deactivated JavaScript (e.g., privacy-savvy users) could be tracked. It is still possible that online services used JavaScript to extract sensitive user information~\cite{senol_unveiling_2023}. Therefore, our study results rather represent a lower baseline of websites that track their users.

We analyzed the HTTP CHs that were sent using the \texttt{Accept-CH} HTTP header. An expired IETF draft suggested that HTTP CHs could also be sent in the TLS 1.3 handshake when using 0-Round-Trip-Time and the TLS Application-Layer Protocol Settings (ALPS) Extension~\cite{davidben-http-client-hint-reliability-03}. However, we found no indications that this was used in practice. We crawled the TLS handshake of the websites and did not see any response where this was included. Furthermore, it shall be noted that also the IETF draft for TLS ALPS has expired~\cite{Benjamin_TLS_2020}%
.

\section{Conclusion}
HTTP CHs were introduced to replace the UAS as a privacy measure in almost all major browsers, but they can be abused to collect even more data from users than was originally possible. In this paper, we present the first long-term study of the use of HTTP CHs in the wild. We found that RBA-instrumented websites tend to collect more detailed user data than those without RBA. Nevertheless, the use of HTTP CHs remains generally low despite their implementation in almost all web browsers. However, in the context of third-party websites, which are often linked to trackers, the usage rate is significantly higher. This is concerning because HTTP CHs allow the retrieval of more data from the client than the UAS provides for, and there are currently no mechanisms in place for users to detect or control this potential data leak. To protect against these practices, browsers should incorporate countermeasures that allow users to control what information they choose to reveal via HTTP CHs, and they should have reasonable default settings that maximize privacy.

\begin{acks}
We would like to thank Rudolf Berrendorf and Javed Razzaq for providing us a huge amount of computational power for our big data analysis. The Platform for Scientific Computing was supported by the \grantsponsor{13FH156IN6}{German Ministry for Education and Research}~, and the \grantsponsor{13FH156IN6}{Ministry for Culture and Science of the state North Rhine-Westphalia}~ (research grant \grantnum{13FH156IN6}{13FH156IN6}).
\end{acks}

\bibliographystyle{ACM-Reference-Format}
\bibliography{bibliography}


\begin{thebibliography}{69}


\ifx \showCODEN    \undefined \def \showCODEN     #1{\unskip}     \fi
\ifx \showDOI      \undefined \def \showDOI       #1{#1}\fi
\ifx \showISBNx    \undefined \def \showISBNx     #1{\unskip}     \fi
\ifx \showISBNxiii \undefined \def \showISBNxiii  #1{\unskip}     \fi
\ifx \showISSN     \undefined \def \showISSN      #1{\unskip}     \fi
\ifx \showLCCN     \undefined \def \showLCCN      #1{\unskip}     \fi
\ifx \shownote     \undefined \def \shownote      #1{#1}          \fi
\ifx \showarticletitle \undefined \def \showarticletitle #1{#1}   \fi
\ifx \showURL      \undefined \def \showURL       {\relax}        \fi
\providecommand\bibfield[2]{#2}
\providecommand\bibinfo[2]{#2}
\providecommand\natexlab[1]{#1}
\providecommand\showeprint[2][]{arXiv:#2}

\bibitem[Acar et~al\mbox{.}(2014)]%
        {acar_the_2014}
\bibfield{author}{\bibinfo{person}{Gunes Acar}, \bibinfo{person}{Christian
  Eubank}, \bibinfo{person}{Steven Englehardt}, \bibinfo{person}{Marc Juarez},
  \bibinfo{person}{Arvind Narayanan}, {and} \bibinfo{person}{Claudia Diaz}.}
  \bibinfo{year}{2014}\natexlab{}.
\newblock \showarticletitle{The Web Never Forgets: Persistent Tracking
  Mechanisms in the Wild}. In \bibinfo{booktitle}{\emph{2014 ACM SIGSAC
  Conference on Computer and Communications Security}} (Scottsdale, Arizona,
  USA) \emph{(\bibinfo{series}{{CCS} '14})}. \bibinfo{publisher}{ACM},
  \bibinfo{pages}{674–689}.
\newblock
\showISBNx{9781450329576}
\urldef\tempurl%
\url{https://doi.org/10.1145/2660267.2660347}
\showDOI{\tempurl}


\bibitem[{Akamai}(2020)]%
        {akamai_loyalty_2020}
\bibfield{author}{\bibinfo{person}{{Akamai}}.} \bibinfo{year}{2020}\natexlab{}.
\newblock \showarticletitle{Loyalty for {Sale} – {Retail} and {Hospitality}
  {Fraud}}.
\newblock \bibinfo{journal}{\emph{[state of the internet] / security}}
  \bibinfo{volume}{6}, \bibinfo{number}{3} (\bibinfo{date}{Oct.}
  \bibinfo{year}{2020}).
\newblock
\urldef\tempurl%
\url{https://web.archive.org/web/20201101013317/https://www.akamai.com/us/en/multimedia/documents/state-of-the-internet/soti-security-loyalty-for-sale-retail-and-hospitality-fraud-report-2020.pdf}
\showURL{%
\tempurl}


\bibitem[Alaca and van Oorschot(2016)]%
        {alaca_device_2016}
\bibfield{author}{\bibinfo{person}{Furkan Alaca} {and} \bibinfo{person}{Paul~C.
  van Oorschot}.} \bibinfo{year}{2016}\natexlab{}.
\newblock \showarticletitle{Device {Fingerprinting} for {Augmenting} {Web}
  {Authentication}: {Classification} and {Analysis} of {Methods}}. In
  \bibinfo{booktitle}{\emph{32nd {Annual} {Computer} {Security} {Applications}
  {Conference}}} (Los Angeles, CA, USA)
  \emph{(\bibinfo{series}{{ACSAC}~’16})}. \bibinfo{publisher}{ACM},
  \bibinfo{pages}{289--301}.
\newblock
\showISBNx{978-1-4503-4771-6}
\urldef\tempurl%
\url{https://doi.org/10.1145/2991079.2991091}
\showDOI{\tempurl}


\bibitem[{Apple}(2024)]%
        {apple_user_2024}
\bibfield{author}{\bibinfo{person}{{Apple}}.} \bibinfo{year}{2024}\natexlab{}.
\newblock \bibinfo{title}{{User Privacy and Data Use}}.
\newblock
\newblock
\urldef\tempurl%
\url{https://web.archive.org/web/20240503143642/https://developer.apple.com/app-store/user-privacy-and-data-use/}
\showURL{%
\tempurl}


\bibitem[{Australian Cyber Security Centre}(2021)]%
        {australian_australian_2021}
\bibfield{author}{\bibinfo{person}{{Australian Cyber Security Centre}}.}
  \bibinfo{year}{2021}\natexlab{}.
\newblock \bibinfo{booktitle}{\emph{Australian {Government} {Information}
  {Security} {Manual}}}.
\newblock \bibinfo{type}{{T}echnical {R}eport}.
\newblock
\urldef\tempurl%
\url{https://web.archive.org/web/20210830131917/https://www.cyber.gov.au/sites/default/files/2021-06/01.\%20ISM\%20-\%20Using\%20the\%20Australian\%20Government\%20Information\%20Security\%20Manual\%20(June\%202021).pdf}
\showURL{%
\tempurl}


\bibitem[Bahrami et~al\mbox{.}(2022)]%
        {bahrami_fp-radar_2022}
\bibfield{author}{\bibinfo{person}{Pouneh~Nikkhah Bahrami},
  \bibinfo{person}{Umar Iqbal}, {and} \bibinfo{person}{Zubair Shafiq}.}
  \bibinfo{year}{2022}\natexlab{}.
\newblock \showarticletitle{{FP}-{Radar}: {Longitudinal} {Measurement} and
  {Early} {Detection} of {Browser} {Fingerprinting}}.
\newblock \bibinfo{journal}{\emph{Proceedings on Privacy Enhancing
  Technologies}} \bibinfo{volume}{2022}, \bibinfo{number}{2}
  (\bibinfo{date}{April} \bibinfo{year}{2022}), \bibinfo{pages}{557--577}.
\newblock
\showISSN{2299-0984}
\urldef\tempurl%
\url{https://doi.org/10.2478/popets-2022-0056}
\showDOI{\tempurl}


\bibitem[Benjamin(2021)]%
        {davidben-http-client-hint-reliability-03}
\bibfield{author}{\bibinfo{person}{David Benjamin}.}
  \bibinfo{year}{2021}\natexlab{}.
\newblock \bibinfo{booktitle}{\emph{{Client Hint Reliability}}}.
\newblock \bibinfo{type}{Internet-Draft}. \bibinfo{institution}{Internet
  Engineering Task Force}.
\newblock
\urldef\tempurl%
\url{https://web.archive.org/web/20230322052424/https://datatracker.ietf.org/doc/draft-davidben-http-client-hint-reliability/03/}
\showURL{%
\tempurl}
\newblock
\shownote{Work in Progress}.


\bibitem[Benjamin and Vasiliev(2020)]%
        {Benjamin_TLS_2020}
\bibfield{author}{\bibinfo{person}{David Benjamin} {and}
  \bibinfo{person}{Victor Vasiliev}.} \bibinfo{year}{2020}\natexlab{}.
\newblock \bibinfo{booktitle}{\emph{{TLS Application-Layer Protocol Settings
  Extension}}}.
\newblock \bibinfo{type}{Internet-Draft}. \bibinfo{institution}{Internet
  Engineering Task Force}.
\newblock
\urldef\tempurl%
\url{https://datatracker.ietf.org/doc/draft-vvv-tls-alps/01/}
\showURL{%
\tempurl}
\newblock
\shownote{Work in Progress}.


\bibitem[Berners-Lee and Connolly(1995)]%
        {berners-lee_hypertext_1995}
\bibfield{author}{\bibinfo{person}{Tim Berners-Lee} {and}
  \bibinfo{person}{Daniel~W. Connolly}.} \bibinfo{year}{1995}\natexlab{}.
\newblock \bibinfo{title}{{Hypertext Markup Language - 2.0}}.
\newblock
\newblock
\urldef\tempurl%
\url{https://doi.org/10.17487/RFC1866}
\showDOI{\tempurl}


\bibitem[Biden~Jr.(2021)]%
        {biden_jr_executive_2021}
\bibfield{author}{\bibinfo{person}{Joseph~R. Biden~Jr.}}
  \bibinfo{year}{2021}\natexlab{}.
\newblock \showarticletitle{Executive {Order} on {Improving} the {Nation}'s
  {Cybersecurity}}.
\newblock \bibinfo{journal}{\emph{The White House}} (\bibinfo{date}{May}
  \bibinfo{year}{2021}).
\newblock
\urldef\tempurl%
\url{https://web.archive.org/web/20220323200621/https://www.whitehouse.gov/briefing-room/presidential-actions/2021/05/12/executive-order-on-improving-the-nations-cybersecurity/}
\showURL{%
\tempurl}


\bibitem[Bujlow et~al\mbox{.}(2017)]%
        {bujilow_a_2017}
\bibfield{author}{\bibinfo{person}{Tomasz Bujlow}, \bibinfo{person}{Valentín
  Carela-Español}, \bibinfo{person}{Josep Solé-Pareta}, {and}
  \bibinfo{person}{Pere Barlet-Ros}.} \bibinfo{year}{2017}\natexlab{}.
\newblock \showarticletitle{A Survey on Web Tracking: Mechanisms, Implications,
  and Defenses}.
\newblock \bibinfo{journal}{\emph{Proc. IEEE}} \bibinfo{volume}{105},
  \bibinfo{number}{8} (\bibinfo{year}{2017}), \bibinfo{pages}{1476--1510}.
\newblock
\urldef\tempurl%
\url{https://doi.org/10.1109/JPROC.2016.2637878}
\showDOI{\tempurl}


\bibitem[{caniuse.com}(2024)]%
        {caniuse_client_2024}
\bibfield{author}{\bibinfo{person}{{caniuse.com}}.}
  \bibinfo{year}{2024}\natexlab{}.
\newblock \bibinfo{title}{{Can I use Client Hint?}}
\newblock
\newblock
\urldef\tempurl%
\url{https://caniuse.com/?search=client+hint}
\showURL{%
\tempurl}


\bibitem[{Chrome Developers}(2023)]%
        {chrome_crux_2023}
\bibfield{author}{\bibinfo{person}{{Chrome Developers}}.}
  \bibinfo{year}{2023}\natexlab{}.
\newblock \bibinfo{title}{{Chrome UX Report}}.
\newblock
\newblock
\urldef\tempurl%
\url{https://web.archive.org/web/20240506101055/https://developer.chrome.com/docs/crux/}
\showURL{%
\tempurl}


\bibitem[{Chrome for Developers}(2020)]%
        {chrome_new_2020}
\bibfield{author}{\bibinfo{person}{{Chrome for Developers}}.}
  \bibinfo{year}{2020}\natexlab{}.
\newblock \bibinfo{title}{{New in Chrome 85}}.
\newblock
\newblock
\urldef\tempurl%
\url{https://web.archive.org/web/20240324075831/https://developer.chrome.com/blog/new-in-chrome-85/}
\showURL{%
\tempurl}


\bibitem[Dao et~al\mbox{.}(2020)]%
        {dao_characterizing_2020}
\bibfield{author}{\bibinfo{person}{Ha Dao}, \bibinfo{person}{Johan Mazel},
  {and} \bibinfo{person}{Kensuke Fukuda}.} \bibinfo{year}{2020}\natexlab{}.
\newblock \showarticletitle{Characterizing {CNAME} {Cloaking}-{Based}
  {Tracking} on the {Web}}. In \bibinfo{booktitle}{\emph{Network {Traffic}
  {Measurement} and {Analysis} {Conference} 2020}} (Berlin, Germany)
  \emph{(\bibinfo{series}{{TMA} '20})}. \bibinfo{publisher}{IFIP Open Digital
  Library}.
\newblock
\urldef\tempurl%
\url{https://web.archive.org/web/20221122113517/https://dl.ifip.org/db/conf/tma/tma2020/tma2020-camera-paper66.pdf}
\showURL{%
\tempurl}


\bibitem[Dao et~al\mbox{.}(2021)]%
        {dao_cname_2021}
\bibfield{author}{\bibinfo{person}{Ha Dao}, \bibinfo{person}{Johan Mazel},
  {and} \bibinfo{person}{Kensuke Fukuda}.} \bibinfo{year}{2021}\natexlab{}.
\newblock \showarticletitle{CNAME Cloaking-Based Tracking on the Web:
  Characterization, Detection, and Protection}.
\newblock \bibinfo{journal}{\emph{IEEE Transactions on Network and Service
  Management}} \bibinfo{volume}{18}, \bibinfo{number}{3}
  (\bibinfo{year}{2021}), \bibinfo{pages}{3873--3888}.
\newblock
\urldef\tempurl%
\url{https://doi.org/10.1109/TNSM.2021.3072874}
\showDOI{\tempurl}


\bibitem[Demir et~al\mbox{.}(2022)]%
        {demir_towards_2022}
\bibfield{author}{\bibinfo{person}{Nurullah Demir}, \bibinfo{person}{Daniel
  Theis}, \bibinfo{person}{Tobias Urban}, {and} \bibinfo{person}{Norbert
  Pohlmann}.} \bibinfo{year}{2022}\natexlab{}.
\newblock \showarticletitle{Towards {Understanding} {First}-{Party} {Cookie}
  {Tracking} in the {Field}}. In \bibinfo{booktitle}{\emph{Sicherheit 2022}}
  \emph{(\bibinfo{series}{Lecture {Notes} in {Informatics}})}.
  \bibinfo{publisher}{Gesellschaft für Informatik}, \bibinfo{pages}{19--34}.
\newblock
\urldef\tempurl%
\url{https://doi.org/10.18420/sicherheit2022_01}
\showDOI{\tempurl}


\bibitem[Doerfler et~al\mbox{.}(2019)]%
        {doerfler_evaluating_2019}
\bibfield{author}{\bibinfo{person}{Periwinkle Doerfler}, \bibinfo{person}{Kurt
  Thomas}, \bibinfo{person}{Maija Marincenko}, \bibinfo{person}{Juri Ranieri},
  \bibinfo{person}{Yu Jiang}, \bibinfo{person}{Angelika Moscicki}, {and}
  \bibinfo{person}{Damon McCoy}.} \bibinfo{year}{2019}\natexlab{}.
\newblock \showarticletitle{Evaluating {Login} {Challenges} As a {Defense}
  {Against} {Account} {Takeover}}. In \bibinfo{booktitle}{\emph{The {World}
  {Wide} {Web} {Conference} 2019}} (San Francisco, CA, USA)
  \emph{(\bibinfo{series}{{WWW}~'19})}. \bibinfo{publisher}{ACM},
  \bibinfo{pages}{372--382}.
\newblock
\showISBNx{978-1-4503-6674-8}
\urldef\tempurl%
\url{https://doi.org/10.1145/3308558.3313481}
\showDOI{\tempurl}


\bibitem[{easylist}(2024)]%
        {easylist_easylist_2024}
\bibfield{author}{\bibinfo{person}{{easylist}}.}
  \bibinfo{year}{2024}\natexlab{}.
\newblock \bibinfo{title}{{EasyList / EasyPrivacy / Fanboy Lists}}.
\newblock
\newblock
\urldef\tempurl%
\url{https://web.archive.org/web/20240427112250/https://github.com/easylist/easylist}
\showURL{%
\tempurl}


\bibitem[Eckersley(2010)]%
        {eckersley_how_2010}
\bibfield{author}{\bibinfo{person}{Peter Eckersley}.}
  \bibinfo{year}{2010}\natexlab{}.
\newblock \showarticletitle{How {Unique} {Is} {Your} {Web} {Browser}?}
\newblock In \bibinfo{booktitle}{\emph{Privacy {Enhancing} {Technologies}}}.
  Vol.~\bibinfo{volume}{6205}. \bibinfo{publisher}{Springer}.
\newblock
\urldef\tempurl%
\url{https://doi.org/10.1007/978-3-642-14527-8_1}
\showDOI{\tempurl}


\bibitem[{European Union}(2016)]%
        {european_union_gdpr_2016}
\bibfield{author}{\bibinfo{person}{{European Union}}.}
  \bibinfo{year}{2016}\natexlab{}.
\newblock \showarticletitle{General {Data} {Protection} {Regulation}}.
\newblock  (\bibinfo{date}{May} \bibinfo{year}{2016}).
\newblock
\urldef\tempurl%
\url{https://web.archive.org/web/20220317074247/https://eur-lex.europa.eu/eli/reg/2016/679/2016-05-04}
\showURL{%
\tempurl}
\newblock
\shownote{{Regulation} (EU) 2016/679}.


\bibitem[Fouad et~al\mbox{.}(2022)]%
        {fouad_my_2022}
\bibfield{author}{\bibinfo{person}{Imane Fouad}, \bibinfo{person}{Cristiana
  Santos}, \bibinfo{person}{Arnaud Legout}, {and} \bibinfo{person}{Nataliia
  Bielova}.} \bibinfo{year}{2022}\natexlab{}.
\newblock \showarticletitle{My {Cookie} is a phoenix: detection, measurement,
  and lawfulness of cookie respawning with browser fingerprinting}.
\newblock \bibinfo{journal}{\emph{Proceedings on Privacy Enhancing
  Technologies}} \bibinfo{volume}{2022}, \bibinfo{number}{3}
  (\bibinfo{date}{July} \bibinfo{year}{2022}), \bibinfo{pages}{79--98}.
\newblock
\showISSN{2299-0984}
\urldef\tempurl%
\url{https://doi.org/10.56553/popets-2022-0063}
\showDOI{\tempurl}


\bibitem[Gavazzi et~al\mbox{.}(2023)]%
        {gavazzi_a_2023}
\bibfield{author}{\bibinfo{person}{Anthony Gavazzi}, \bibinfo{person}{Ryan
  Williams}, \bibinfo{person}{Engin Kirda}, \bibinfo{person}{Long Lu},
  \bibinfo{person}{Andre King}, \bibinfo{person}{Andy Davis}, {and}
  \bibinfo{person}{Tim Leek}.} \bibinfo{year}{2023}\natexlab{}.
\newblock \showarticletitle{A Study of Multi-Factor and Risk-Based
  Authentication Availability}. In \bibinfo{booktitle}{\emph{32nd USENIX
  Security Symposium}} (Anaheim, CA, USA) \emph{(\bibinfo{series}{{USENIX}
  Security '23})}. \bibinfo{publisher}{USENIX Association},
  \bibinfo{pages}{2043--2060}.
\newblock
\urldef\tempurl%
\url{https://www.usenix.org/conference/usenixsecurity23/presentation/gavazzi}
\showURL{%
\tempurl}


\bibitem[{Google Cloud}(2024)]%
        {google_bigquery_2024}
\bibfield{author}{\bibinfo{person}{{Google Cloud}}.}
  \bibinfo{year}{2024}\natexlab{}.
\newblock \bibinfo{title}{{BigQuery}}.
\newblock
\newblock
\urldef\tempurl%
\url{https://web.archive.org/web/20240504103158/https://cloud.google.com/bigquery?hl=en}
\showURL{%
\tempurl}


\bibitem[Grassi et~al\mbox{.}(2017)]%
        {grassi_digital_2017}
\bibfield{author}{\bibinfo{person}{Paul~A. Grassi}, \bibinfo{person}{James~L.
  Fenton}, \bibinfo{person}{Elaine~M. Newton}, \bibinfo{person}{Ray~A.
  Perlner}, \bibinfo{person}{Andrew~R. Regenscheid},
  \bibinfo{person}{William~E. Burr}, \bibinfo{person}{Justin~P. Richer},
  \bibinfo{person}{Naomi~B. Lefkovitz}, \bibinfo{person}{Jamie~M. Danker},
  \bibinfo{person}{Yee-Yin Choong}, \bibinfo{person}{Kristen~K. Greene}, {and}
  \bibinfo{person}{Mary~F. Theofanos}.} \bibinfo{year}{2017}\natexlab{}.
\newblock \bibinfo{booktitle}{\emph{{Digital Identity Guidelines:
  Authentication and Lifecycle Management}}}.
\newblock \bibinfo{type}{{T}echnical {R}eport} NIST SP 800-63b.
  \bibinfo{institution}{National Institute of Standards and Technology}.
\newblock
\urldef\tempurl%
\url{https://doi.org/10.6028/NIST.SP.800-63b}
\showDOI{\tempurl}


\bibitem[Grigorik(2013)]%
        {grigorik-http-client-hints-00}
\bibfield{author}{\bibinfo{person}{Ilya Grigorik}.}
  \bibinfo{year}{2013}\natexlab{}.
\newblock \bibinfo{booktitle}{\emph{{HTTP Client Hints}}}.
\newblock \bibinfo{type}{Internet-Draft}. \bibinfo{institution}{Internet
  Engineering Task Force}.
\newblock
\urldef\tempurl%
\url{https://web.archive.org/web/20230528025111/https://datatracker.ietf.org/doc/draft-grigorik-http-client-hints/00/}
\showURL{%
\tempurl}
\newblock
\shownote{Work in Progress}.


\bibitem[Grigorik(2015)]%
        {ietf-httpbis-client-hints-00}
\bibfield{author}{\bibinfo{person}{Ilya Grigorik}.}
  \bibinfo{year}{2015}\natexlab{}.
\newblock \bibinfo{booktitle}{\emph{{HTTP Client Hints}}}.
\newblock \bibinfo{type}{Internet-Draft}. \bibinfo{institution}{Internet
  Engineering Task Force}.
\newblock
\urldef\tempurl%
\url{https://web.archive.org/web/20231004041902/https://datatracker.ietf.org/doc/draft-ietf-httpbis-client-hints/00/}
\showURL{%
\tempurl}
\newblock
\shownote{Work in Progress}.


\bibitem[Grigorik and Weiss(2021a)]%
        {rfc8942}
\bibfield{author}{\bibinfo{person}{Ilya Grigorik} {and} \bibinfo{person}{Yoav
  Weiss}.} \bibinfo{year}{2021}\natexlab{a}.
\newblock \bibinfo{title}{{HTTP Client Hints}}.
\newblock \bibinfo{howpublished}{RFC 8942}.
\newblock
\urldef\tempurl%
\url{https://doi.org/10.17487/RFC8942}
\showDOI{\tempurl}


\bibitem[Grigorik and Weiss(2021b)]%
        {grigorik_http_2021}
\bibfield{author}{\bibinfo{person}{I. Grigorik} {and} \bibinfo{person}{Y.
  Weiss}.} \bibinfo{year}{2021}\natexlab{b}.
\newblock \bibinfo{booktitle}{\emph{{HTTP} {Client} {Hints}}}.
\newblock \bibinfo{type}{{T}echnical {R}eport} RFC8942.
\newblock
\urldef\tempurl%
\url{https://doi.org/10.17487/RFC8942}
\showDOI{\tempurl}


\bibitem[Haber(2020)]%
        {haber_attack_2020}
\bibfield{author}{\bibinfo{person}{Morey~J. Haber}.}
  \bibinfo{year}{2020}\natexlab{}.
\newblock \showarticletitle{Attack {Vectors}}.
\newblock In \bibinfo{booktitle}{\emph{Privileged {Attack} {Vectors}:
  {Building} {Effective} {Cyber}-{Defense} {Strategies} to {Protect}
  {Organizations}}}. \bibinfo{publisher}{Apress}, \bibinfo{address}{Berkeley,
  CA, USA}, \bibinfo{pages}{65--85}.
\newblock
\showISBNx{978-1-4842-5914-6}
\urldef\tempurl%
\url{https://doi.org/10.1007/978-1-4842-5914-6_4}
\showDOI{\tempurl}


\bibitem[Han and Lee(2023)]%
        {han_detecting_2023}
\bibfield{author}{\bibinfo{person}{Alex~Heunhe Han} {and}
  \bibinfo{person}{Dong~Hoon Lee}.} \bibinfo{year}{2023}\natexlab{}.
\newblock \showarticletitle{Detecting Risky Authentication Using the OpenID
  Connect Token Exchange Time}.
\newblock \bibinfo{journal}{\emph{Sensors}} \bibinfo{volume}{23},
  \bibinfo{number}{19} (\bibinfo{year}{2023}).
\newblock
\showISSN{1424-8220}
\urldef\tempurl%
\url{https://doi.org/10.3390/s23198256}
\showDOI{\tempurl}


\bibitem[Hantke et~al\mbox{.}(2023)]%
        {hantke_you_2023}
\bibfield{author}{\bibinfo{person}{Florian Hantke}, \bibinfo{person}{Stefano
  Calzavara}, \bibinfo{person}{Moritz Wilhelm}, \bibinfo{person}{Alvise
  Rabitti}, {and} \bibinfo{person}{Ben Stock}.}
  \bibinfo{year}{2023}\natexlab{}.
\newblock \showarticletitle{You {Call} {This} {Archaeology}? {Evaluating} {Web}
  {Archives} for {Reproducible} {Web} {Security} {Measurements}}. In
  \bibinfo{booktitle}{\emph{2023 {ACM} {SIGSAC} {Conference} on {Computer} and
  {Communications} {Security}}} (Copenhagen, Denmark)
  \emph{(\bibinfo{series}{{CCS} '23})}. \bibinfo{publisher}{ACM},
  \bibinfo{pages}{3168--3182}.
\newblock
\urldef\tempurl%
\url{https://doi.org/10.1145/3576915.3616688}
\showDOI{\tempurl}


\bibitem[{HTTP Archive}(2023)]%
        {httparchive_faq_2023}
\bibfield{author}{\bibinfo{person}{{HTTP Archive}}.}
  \bibinfo{year}{2023}\natexlab{}.
\newblock \bibinfo{title}{{Frequently Asked Questions}}.
\newblock
\newblock
\urldef\tempurl%
\url{https://web.archive.org/web/20240506061519/https://httparchive.org/faq}
\showURL{%
\tempurl}


\bibitem[{Internet Archive}(2023)]%
        {wayback_machine_2023}
\bibfield{author}{\bibinfo{person}{{Internet Archive}}.}
  \bibinfo{year}{2023}\natexlab{}.
\newblock \bibinfo{title}{{Wayback Machine}}.
\newblock
\newblock
\urldef\tempurl%
\url{https://web.archive.org}
\showURL{%
\tempurl}


\bibitem[Intumwayase et~al\mbox{.}(2023)]%
        {intumwayase_ua-radar_2023}
\bibfield{author}{\bibinfo{person}{Jean~Luc Intumwayase},
  \bibinfo{person}{Imane Fouad}, \bibinfo{person}{Pierre Laperdrix}, {and}
  \bibinfo{person}{Romain Rouvoy}.} \bibinfo{year}{2023}\natexlab{}.
\newblock \showarticletitle{{UA}-{Radar}: {Exploring} the {Impact} of {User}
  {Agents} on the {Web}}. In \bibinfo{booktitle}{\emph{22nd {Workshop} on
  {Privacy} in the {Electronic} {Society}}} (Copenhagen, Denmark)
  \emph{(\bibinfo{series}{{WPES} '23})}. \bibinfo{publisher}{ACM},
  \bibinfo{pages}{31--43}.
\newblock
\urldef\tempurl%
\url{https://doi.org/10.1145/3603216.3624958}
\showDOI{\tempurl}


\bibitem[Iqbal et~al\mbox{.}(2021)]%
        {iqbal_fingerprinting_2021}
\bibfield{author}{\bibinfo{person}{Umar Iqbal}, \bibinfo{person}{Steven
  Englehardt}, {and} \bibinfo{person}{Zubair Shafiq}.}
  \bibinfo{year}{2021}\natexlab{}.
\newblock \showarticletitle{Fingerprinting the {Fingerprinters}: {Learning} to
  {Detect} {Browser} {Fingerprinting} {Behaviors}}. In
  \bibinfo{booktitle}{\emph{2021 {IEEE} {Symposium} on {Security} and
  {Privacy}}} (San Francisco, CA, USA) \emph{(\bibinfo{series}{{SP} '21})}.
  \bibinfo{publisher}{IEEE}, \bibinfo{pages}{1143--1161}.
\newblock
\urldef\tempurl%
\url{https://doi.org/10.1109/SP40001.2021.00017}
\showDOI{\tempurl}


\bibitem[Jonker et~al\mbox{.}(2019)]%
        {jonker_fingerprint_2019}
\bibfield{author}{\bibinfo{person}{Hugo Jonker}, \bibinfo{person}{Benjamin
  Krumnow}, {and} \bibinfo{person}{Gabry Vlot}.}
  \bibinfo{year}{2019}\natexlab{}.
\newblock \showarticletitle{Fingerprint Surface-Based Detection of Web Bot
  Detectors}. In \bibinfo{booktitle}{\emph{ESORICS '19}}.
  \bibinfo{publisher}{Springer}, \bibinfo{pages}{586--605}.
\newblock
\urldef\tempurl%
\url{https://doi.org/10.1007/978-3-030-29962-0_28}
\showDOI{\tempurl}


\bibitem[Laperdrix et~al\mbox{.}(2020)]%
        {laperdrix_browser_2020}
\bibfield{author}{\bibinfo{person}{Pierre Laperdrix}, \bibinfo{person}{Nataliia
  Bielova}, \bibinfo{person}{Benoit Baudry}, {and} \bibinfo{person}{Gildas
  Avoine}.} \bibinfo{year}{2020}\natexlab{}.
\newblock \showarticletitle{Browser Fingerprinting: A Survey}.
\newblock \bibinfo{journal}{\emph{ACM Trans. Web}} \bibinfo{volume}{14},
  \bibinfo{number}{2}, Article \bibinfo{articleno}{8} (\bibinfo{date}{apr}
  \bibinfo{year}{2020}), \bibinfo{numpages}{33}~pages.
\newblock
\showISSN{1559-1131}
\urldef\tempurl%
\url{https://doi.org/10.1145/3386040}
\showDOI{\tempurl}


\bibitem[Le et~al\mbox{.}(2021)]%
        {le_cv-inspector_2021}
\bibfield{author}{\bibinfo{person}{Hieu Le}, \bibinfo{person}{Athina
  Markopoulou}, {and} \bibinfo{person}{Zubair Shafiq}.}
  \bibinfo{year}{2021}\natexlab{}.
\newblock \showarticletitle{{CV}-{Inspector}: {Towards} {Automating}
  {Detection} of {Adblock} {Circumvention}}. In \bibinfo{booktitle}{\emph{2021
  {Network} and {Distributed} {System} {Security} {Symposium}}} (Virtual)
  \emph{(\bibinfo{series}{{NDSS} '21})}. \bibinfo{publisher}{Internet Society}.
\newblock
\showISBNx{978-1-891562-66-2}
\urldef\tempurl%
\url{https://doi.org/10.14722/ndss.2021.24055}
\showDOI{\tempurl}


\bibitem[Leith(2021)]%
        {leith_web_2021}
\bibfield{author}{\bibinfo{person}{Douglas~J. Leith}.}
  \bibinfo{year}{2021}\natexlab{}.
\newblock \showarticletitle{Web {Browser} {Privacy}: {What} {Do} {Browsers}
  {Say} {When} {They} {Phone} {Home}?}
\newblock \bibinfo{journal}{\emph{IEEE Access}}  \bibinfo{volume}{9}
  (\bibinfo{year}{2021}), \bibinfo{pages}{41615--41627}.
\newblock
\showISSN{2169-3536}
\urldef\tempurl%
\url{https://doi.org/10.1109/ACCESS.2021.3065243}
\showDOI{\tempurl}


\bibitem[Makowski and P\"{o}hn(2023)]%
        {makowski_evaluation_2023}
\bibfield{author}{\bibinfo{person}{Jan-Phillip Makowski} {and}
  \bibinfo{person}{Daniela P\"{o}hn}.} \bibinfo{year}{2023}\natexlab{}.
\newblock \showarticletitle{Evaluation of Real-World Risk-Based Authentication
  at Online Services Revisited: Complexity Wins}. In
  \bibinfo{booktitle}{\emph{18th {International Workshop on Frontiers in
  Availability, Reliability and Security}}} (Benevento, Italy)
  \emph{(\bibinfo{series}{{FARES} '23})}. \bibinfo{publisher}{ACM}.
\newblock
\urldef\tempurl%
\url{https://doi.org/10.1145/3600160.3605024}
\showDOI{\tempurl}


\bibitem[{McAfee}(2023)]%
        {mcafee_check_2023}
\bibfield{author}{\bibinfo{person}{{McAfee}}.} \bibinfo{year}{2023}\natexlab{}.
\newblock \bibinfo{title}{{Check Single URL}}.
\newblock
\newblock
\urldef\tempurl%
\url{https://sitelookup.mcafee.com/}
\showURL{%
\tempurl}


\bibitem[{mdn web docs}(2024)]%
        {mdn_user_2024}
\bibfield{author}{\bibinfo{person}{{mdn web docs}}.}
  \bibinfo{year}{2024}\natexlab{}.
\newblock \bibinfo{title}{{User-Agent}}.
\newblock
\newblock
\urldef\tempurl%
\url{https://web.archive.org/web/20240224150202/https://developer.mozilla.org/en-US/docs/Web/HTTP/Headers/User-Agent}
\showURL{%
\tempurl}


\bibitem[Merewood and Weiss(2020)]%
        {merewood_improving_2020}
\bibfield{author}{\bibinfo{person}{Rowan Merewood} {and} \bibinfo{person}{Yoav
  Weiss}.} \bibinfo{year}{2020}\natexlab{}.
\newblock \bibinfo{title}{{Improving user privacy and developer experience with
  User-Agent Client Hints}}.
\newblock
\newblock
\urldef\tempurl%
\url{https://web.archive.org/web/20221208053222/https://developer.chrome.com/articles/user-agent-client-hints/}
\showURL{%
\tempurl}


\bibitem[{National Cyber Security Centre}(2022)]%
        {national_cyber_security_centre_cloud_2018}
\bibfield{author}{\bibinfo{person}{{National Cyber Security Centre}}.}
  \bibinfo{year}{2022}\natexlab{}.
\newblock \bibinfo{booktitle}{\emph{{Cloud Security Guidance: Principle 10:
  Identity and Authentication}}}.
\newblock \bibinfo{type}{{T}echnical {R}eport}.
\newblock
\urldef\tempurl%
\url{https://web.archive.org/web/20220518230012/https://www.ncsc.gov.uk/collection/cloud/the-cloud-security-principles/principle-10-identity-and-authentication}
\showURL{%
\tempurl}


\bibitem[Ney(2022)]%
        {ney_prepare_2022}
\bibfield{author}{\bibinfo{person}{Heremy Ney}.}
  \bibinfo{year}{2022}\natexlab{}.
\newblock \bibinfo{title}{{Prepare for User-Agent Reduction changes in
  October}}.
\newblock
\newblock
\urldef\tempurl%
\url{https://web.archive.org/web/20221003134709/https://developer.chrome.com/blog/user-agent-reduction-oct-2022-updates/}
\showURL{%
\tempurl}


\bibitem[Nielsen et~al\mbox{.}(1996)]%
        {nielsen_http_1996}
\bibfield{author}{\bibinfo{person}{Henrik Nielsen}, \bibinfo{person}{Roy~T.
  Fielding}, {and} \bibinfo{person}{Tim Berners-Lee}.}
  \bibinfo{year}{1996}\natexlab{}.
\newblock \bibinfo{title}{{Hypertext Transfer Protocol -- HTTP/1.0}}.
\newblock
\newblock
\urldef\tempurl%
\url{https://doi.org/10.17487/RFC1945}
\showDOI{\tempurl}


\bibitem[Papadogiannakis et~al\mbox{.}(2021)]%
        {papadogiannakis_user_2021}
\bibfield{author}{\bibinfo{person}{Emmanouil Papadogiannakis},
  \bibinfo{person}{Panagiotis Papadopoulos}, \bibinfo{person}{Nicolas
  Kourtellis}, {and} \bibinfo{person}{Evangelos~P. Markatos}.}
  \bibinfo{year}{2021}\natexlab{}.
\newblock \showarticletitle{User {Tracking} in the {Post}-cookie {Era}: {How}
  {Websites} {Bypass} {GDPR} {Consent} to {Track} {Users}}. In
  \bibinfo{booktitle}{\emph{The {Web} {Conference} 2021}} (Ljubljana, Slovenia)
  \emph{(\bibinfo{series}{{WWW} '21})}. \bibinfo{publisher}{ACM},
  \bibinfo{pages}{2130--2141}.
\newblock
\showISBNx{978-1-4503-8312-7}
\urldef\tempurl%
\url{https://doi.org/10.1145/3442381.3450056}
\showDOI{\tempurl}


\bibitem[Pugliese et~al\mbox{.}(2020)]%
        {pugliese_long-term_2020}
\bibfield{author}{\bibinfo{person}{Gaston Pugliese}, \bibinfo{person}{Christian
  Riess}, \bibinfo{person}{Freya Gassmann}, {and} \bibinfo{person}{Zinaida
  Benenson}.} \bibinfo{year}{2020}\natexlab{}.
\newblock \showarticletitle{Long-{Term} {Observation} on {Browser}
  {Fingerprinting}: {Users}’ {Trackability} and {Perspective}}.
\newblock \bibinfo{journal}{\emph{Proceedings on Privacy Enhancing
  Technologies}} \bibinfo{volume}{2020}, \bibinfo{number}{2}
  (\bibinfo{date}{April} \bibinfo{year}{2020}), \bibinfo{pages}{558--577}.
\newblock
\showISSN{2299-0984}
\urldef\tempurl%
\url{https://doi.org/10.2478/popets-2020-0041}
\showDOI{\tempurl}


\bibitem[Sanchez-Rola et~al\mbox{.}(2019)]%
        {sanchez-rola_can_2019}
\bibfield{author}{\bibinfo{person}{Iskander Sanchez-Rola},
  \bibinfo{person}{Matteo Dell'Amico}, \bibinfo{person}{Platon Kotzias},
  \bibinfo{person}{Davide Balzarotti}, \bibinfo{person}{Leyla Bilge},
  \bibinfo{person}{Pierre-Antoine Vervier}, {and} \bibinfo{person}{Igor
  Santos}.} \bibinfo{year}{2019}\natexlab{}.
\newblock \showarticletitle{Can {I} {Opt} {Out} {Yet}?: {GDPR} and the {Global}
  {Illusion} of {Cookie} {Control}}. In \bibinfo{booktitle}{\emph{2019 {ACM}
  {Asia} {Conference} on {Computer} and {Communications} {Security}}}
  (Auckland, New Zealand) \emph{(\bibinfo{series}{{AsiaCCS} '19})}.
  \bibinfo{publisher}{ACM}, \bibinfo{pages}{340--351}.
\newblock
\urldef\tempurl%
\url{https://doi.org/10.1145/3321705.3329806}
\showDOI{\tempurl}


\bibitem[Sanchez-Rola et~al\mbox{.}(2017)]%
        {sanchez-rola_web_2017}
\bibfield{author}{\bibinfo{person}{Iskander Sanchez-Rola},
  \bibinfo{person}{Xabier Ugarte-Pedrero}, \bibinfo{person}{Igor Santos}, {and}
  \bibinfo{person}{Pablo~G. Bringas}.} \bibinfo{year}{2017}\natexlab{}.
\newblock \showarticletitle{The web is watching you: {A} comprehensive review
  of web-tracking techniques and countermeasures}.
\newblock \bibinfo{journal}{\emph{Logic Journal of IGPL}} \bibinfo{volume}{25},
  \bibinfo{number}{1} (\bibinfo{date}{Feb.} \bibinfo{year}{2017}),
  \bibinfo{pages}{18--29}.
\newblock
\urldef\tempurl%
\url{https://doi.org/10.1093/jigpal/jzw041}
\showDOI{\tempurl}


\bibitem[Schöni et~al\mbox{.}(2024)]%
        {schoni_block_2024}
\bibfield{author}{\bibinfo{person}{Lorin Schöni}, \bibinfo{person}{Karel
  Kubicek}, {and} \bibinfo{person}{Verena Zimmermann}.}
  \bibinfo{year}{2024}\natexlab{}.
\newblock \showarticletitle{Block {Cookies}, {Not} {Websites}: {Analysing}
  {Mental} {Models} and {Usability} of the {Privacy}-{Preserving} {Browser}
  {Extension} {CookieBlock}}.
\newblock \bibinfo{journal}{\emph{Proceedings on Privacy Enhancing
  Technologies}} \bibinfo{volume}{2024}, \bibinfo{number}{1}
  (\bibinfo{date}{Jan.} \bibinfo{year}{2024}), \bibinfo{pages}{192--216}.
\newblock
\urldef\tempurl%
\url{https://doi.org/10.56553/popets-2024-0012}
\showDOI{\tempurl}


\bibitem[Senol and Acar(2023)]%
        {senol_unveiling_2023}
\bibfield{author}{\bibinfo{person}{Asuman Senol} {and} \bibinfo{person}{Gunes
  Acar}.} \bibinfo{year}{2023}\natexlab{}.
\newblock \showarticletitle{Unveiling the {Impact} of {User}-{Agent}
  {Reduction} and {Client} {Hints}: {A} {Measurement} {Study}}. In
  \bibinfo{booktitle}{\emph{22nd {Workshop} on {Privacy} in the {Electronic}
  {Society}}} (Copenhagen, Denmark) \emph{(\bibinfo{series}{{WPES} '23})}.
  \bibinfo{publisher}{ACM}, \bibinfo{pages}{91--106}.
\newblock
\urldef\tempurl%
\url{https://doi.org/10.1145/3603216.3624965}
\showDOI{\tempurl}


\bibitem[Solomos et~al\mbox{.}(2020)]%
        {solomos_clash_2020}
\bibfield{author}{\bibinfo{person}{Konstantinos Solomos},
  \bibinfo{person}{Panagiotis Ilia}, \bibinfo{person}{Sotiris Ioannidis}, {and}
  \bibinfo{person}{Nicolas Kourtellis}.} \bibinfo{year}{2020}\natexlab{}.
\newblock \showarticletitle{Clash of the {Trackers}: {Measuring} the
  {Evolution} of the {Online} {Tracking} {Ecosystem}}. In
  \bibinfo{booktitle}{\emph{Network {Traffic} {Measurement} and {Analysis}
  {Conference} 2020}} (Berlin, Germany). \bibinfo{publisher}{IFIP Open Digital
  Library}.
\newblock
\urldef\tempurl%
\url{https://web.archive.org/web/20201127171719/https://tma.ifip.org/2020/wp-content/uploads/sites/9/2020/06/tma2020-camera-paper36.pdf}
\showURL{%
\tempurl}


\bibitem[Solomos et~al\mbox{.}(2021)]%
        {solomos_tales_2021}
\bibfield{author}{\bibinfo{person}{Konstantinos Solomos}, \bibinfo{person}{John
  Kristoff}, \bibinfo{person}{Chris Kanich}, {and} \bibinfo{person}{Jason
  Polakis}.} \bibinfo{year}{2021}\natexlab{}.
\newblock \showarticletitle{Tales of favicons and caches: Persistent tracking
  in modern browsers}. In \bibinfo{booktitle}{\emph{Network and Distributed
  System Security Symposium}} \emph{(\bibinfo{series}{{NDSS} '21})}.
\newblock
\urldef\tempurl%
\url{https://doi.org/10.14722/ndss.2021.24202}
\showDOI{\tempurl}


\bibitem[{State of California}(2018)]%
        {california_ccpa_2018}
\bibfield{author}{\bibinfo{person}{{State of California}}.}
  \bibinfo{year}{2018}\natexlab{}.
\newblock \showarticletitle{{California} {Consumer} {Privacy} {Act}}.
\newblock  (\bibinfo{date}{June} \bibinfo{year}{2018}).
\newblock
\urldef\tempurl%
\url{https://web.archive.org/web/20220323195500if_/https://leginfo.legislature.ca.gov/faces/billTextClient.xhtml?bill_id=201720180AB375}
\showURL{%
\tempurl}
\newblock
\shownote{{Assembly} {Bill} {No.} 375}.


\bibitem[{Statscounter}(2024)]%
        {statscounter_market_2024}
\bibfield{author}{\bibinfo{person}{{Statscounter}}.}
  \bibinfo{year}{2024}\natexlab{}.
\newblock \bibinfo{title}{{Browser Market Share Worldwide}}.
\newblock
\newblock
\urldef\tempurl%
\url{https://gs.statcounter.com/}
\showURL{%
\tempurl}


\bibitem[{The Chromium Projects}(2021)]%
        {chromium_user_2021}
\bibfield{author}{\bibinfo{person}{{The Chromium Projects}}.}
  \bibinfo{year}{2021}\natexlab{}.
\newblock \bibinfo{title}{{User-Agent Client-Hints?}}
\newblock
\newblock
\urldef\tempurl%
\url{https://web.archive.org/web/20240000000000*/https://www.chromium.org/updates/ua-ch}
\showURL{%
\tempurl}


\bibitem[{The Chromium Projects}(2023)]%
        {chromium_user_2023}
\bibfield{author}{\bibinfo{person}{{The Chromium Projects}}.}
  \bibinfo{year}{2023}\natexlab{}.
\newblock \bibinfo{title}{{User-Agent Reduction?}}
\newblock
\newblock
\urldef\tempurl%
\url{https://web.archive.org/web/20240412234718/https://www.chromium.org/updates/ua-reduction/}
\showURL{%
\tempurl}


\bibitem[Unsel et~al\mbox{.}(2023)]%
        {Unsel_Risk_2023}
\bibfield{author}{\bibinfo{person}{Vincent Unsel}, \bibinfo{person}{Stephan
  Wiefling}, \bibinfo{person}{Nils Gruschka}, {and} \bibinfo{person}{Luigi
  Lo~Iacono}.} \bibinfo{year}{2023}\natexlab{}.
\newblock \showarticletitle{{Risk-Based Authentication for OpenStack: A Fully
  Functional Implementation and Guiding Example}}. In
  \bibinfo{booktitle}{\emph{{13th ACM Conference on Data and Application
  Security and Privacy}}} (Charlotte, NC, USA)
  \emph{(\bibinfo{series}{{CODASPY} '23})}. \bibinfo{publisher}{ACM},
  \bibinfo{pages}{237–--243}.
\newblock
\urldef\tempurl%
\url{https://doi.org/10.1145/3577923.3583634}
\showDOI{\tempurl}


\bibitem[Vastel et~al\mbox{.}(2020)]%
        {vastel_fp-crawlers_2020}
\bibfield{author}{\bibinfo{person}{Antoine Vastel}, \bibinfo{person}{Walter
  Rudametkin}, \bibinfo{person}{Romain Rouvoy}, {and} \bibinfo{person}{Xavier
  Blanc}.} \bibinfo{year}{2020}\natexlab{}.
\newblock \showarticletitle{{FP}-{Crawlers}: {Studying} the {Resilience} of
  {Browser} {Fingerprinting} to {Block} {Crawlers}}. In
  \bibinfo{booktitle}{\emph{Proceedings 2020 {Workshop} on {Measurements},
  {Attacks}, and {Defenses} for the {Web}}} (San Diego, CA).
  \bibinfo{publisher}{Internet Society}.
\newblock
\showISBNx{978-1-891562-63-1}
\urldef\tempurl%
\url{https://doi.org/10.14722/madweb.2020.23010}
\showDOI{\tempurl}


\bibitem[Weiss(2023)]%
        {weiss_client_2023}
\bibfield{author}{\bibinfo{person}{Yoav Weiss}.}
  \bibinfo{year}{2023}\natexlab{}.
\newblock \bibinfo{booktitle}{\emph{{Client Hints Infrastructure}}}.
\newblock \bibinfo{type}{{T}echnical {R}eport}.
\newblock
\urldef\tempurl%
\url{https://wicg.github.io/client-hints-infrastructure/}
\showURL{%
\tempurl}
\newblock
\shownote{Draft Community Group Report, 14 July 2023}.


\bibitem[Wiefling et~al\mbox{.}(2021a)]%
        {Wiefling_Whats_2021}
\bibfield{author}{\bibinfo{person}{Stephan Wiefling}, \bibinfo{person}{Markus
  D\"{u}rmuth}, {and} \bibinfo{person}{Luigi Lo~Iacono}.}
  \bibinfo{year}{2021}\natexlab{a}.
\newblock \showarticletitle{What’s in {Score} for {Website} {Users}: {A}
  {Data}-{Driven} {Long}-{Term} {Study} on {Risk}-{Based} {Authentication}
  {Characteristics}}. In \bibinfo{booktitle}{\emph{25th {International}
  {Conference} on {Financial} {Cryptography} and {Data} {Security}}} (Grenada)
  \emph{(\bibinfo{series}{{FC} '21})}. \bibinfo{publisher}{Springer},
  \bibinfo{pages}{361--381}.
\newblock
\urldef\tempurl%
\url{https://doi.org/10.1007/978-3-662-64331-0_19}
\showDOI{\tempurl}


\bibitem[Wiefling et~al\mbox{.}(2019a)]%
        {wiefling_even_2019}
\bibfield{author}{\bibinfo{person}{Stephan Wiefling}, \bibinfo{person}{Nils
  Gruschka}, {and} \bibinfo{person}{Luigi Lo~Iacono}.}
  \bibinfo{year}{2019}\natexlab{a}.
\newblock \showarticletitle{Even {Turing} {Should} {Sometimes} {Not} {Be}
  {Able} {To} {Tell}: {Mimicking} {Humanoid} {Usage} {Behavior} for
  {Exploratory} {Studies} of {Online} {Services}}. In
  \bibinfo{booktitle}{\emph{24th {Nordic} {Conference} on {Secure} {IT}
  {Systems}}} (Aalborg, Denmark) \emph{(\bibinfo{series}{{NordSec '19}})}.
  \bibinfo{publisher}{Springer}, \bibinfo{pages}{188--203}.
\newblock
\urldef\tempurl%
\url{https://doi.org/10.1007/978-3-030-35055-0_12}
\showDOI{\tempurl}


\bibitem[Wiefling et~al\mbox{.}(2023a)]%
        {Wiefling_Pump_2022}
\bibfield{author}{\bibinfo{person}{Stephan Wiefling},
  \bibinfo{person}{Paul~René Jørgensen}, \bibinfo{person}{Sigurd Thunem},
  {and} \bibinfo{person}{Luigi {Lo Iacono}}.} \bibinfo{year}{2023}\natexlab{a}.
\newblock \showarticletitle{Pump {Up} {Password} {Security}! {Evaluating} and
  {Enhancing} {Risk}-{Based} {Authentication} on a {Real}-{World}
  {Large}-{Scale} {Online} {Service}}.
\newblock \bibinfo{journal}{\emph{{ACM} {Transactions} on {Privacy} and
  {Security}}} \bibinfo{volume}{26}, \bibinfo{number}{1}, Article
  \bibinfo{articleno}{6} (\bibinfo{date}{Feb.} \bibinfo{year}{2023}).
\newblock
\showISSN{2471-2566}
\urldef\tempurl%
\url{https://doi.org/10.1145/3546069}
\showDOI{\tempurl}


\bibitem[Wiefling et~al\mbox{.}(2019b)]%
        {wiefling_is_2019}
\bibfield{author}{\bibinfo{person}{Stephan Wiefling}, \bibinfo{person}{Luigi
  Lo~Iacono}, {and} \bibinfo{person}{Markus D\"urmuth}.}
  \bibinfo{year}{2019}\natexlab{b}.
\newblock \showarticletitle{Is {This} {Really} {You}? {An} {Empirical} {Study}
  on {Risk}-{Based} {Authentication} {Applied} in the {Wild}}. In
  \bibinfo{booktitle}{\emph{34th {IFIP} {TC}-11 {International} {Conference} on
  {Information} {Security} and {Privacy} {Protection}}} (Lisbon, Portugal)
  \emph{(\bibinfo{series}{{IFIP} {SEC} '19})}. \bibinfo{publisher}{Springer},
  \bibinfo{pages}{134--148}.
\newblock
\urldef\tempurl%
\url{https://doi.org/10.1007/978-3-030-22312-0\_10}
\showDOI{\tempurl}


\bibitem[Wiefling et~al\mbox{.}(2021b)]%
        {wiefling_privacy_2021}
\bibfield{author}{\bibinfo{person}{Stephan Wiefling}, \bibinfo{person}{Jan
  Tolsdorf}, {and} \bibinfo{person}{Luigi Lo~Iacono}.}
  \bibinfo{year}{2021}\natexlab{b}.
\newblock \showarticletitle{{Privacy} {Considerations} for {Risk}-{Based}
  {Authentication} {Systems}}. In \bibinfo{booktitle}{\emph{2021
  {International} {Workshop} on {Privacy} {Engineering}}} (Vienna, Austria)
  \emph{(\bibinfo{series}{{IWPE}~'21})}. \bibinfo{publisher}{{IEEE}},
  \bibinfo{pages}{320--327}.
\newblock
\urldef\tempurl%
\url{https://doi.org/10.1109/EuroSPW54576.2021.00040}
\showDOI{\tempurl}


\bibitem[Wiefling et~al\mbox{.}(2023b)]%
        {wiefling_data_2022}
\bibfield{author}{\bibinfo{person}{Stephan Wiefling}, \bibinfo{person}{Jan
  Tolsdorf}, {and} \bibinfo{person}{Luigi Lo~Iacono}.}
  \bibinfo{year}{2023}\natexlab{b}.
\newblock \showarticletitle{Data {Protection} {Officers}' {Perspectives} on
  {Privacy} {Challenges} in {Digital} {Ecosystems}}. In
  \bibinfo{booktitle}{\emph{4th {Workshop} on {Security}, {Privacy},
  {Organizations}, and {Systems} {Engineering}}} (Copenhagen, Denmark)
  \emph{(\bibinfo{series}{{SPOSE} '22})}. \bibinfo{publisher}{Springer}.
\newblock
\urldef\tempurl%
\url{https://doi.org/10.1007/978-3-031-25460-4_13}
\showDOI{\tempurl}


\bibitem[Zuboff(2019)]%
        {zuboff_surveillance_2019}
\bibfield{author}{\bibinfo{person}{Shoshana Zuboff}.}
  \bibinfo{year}{2019}\natexlab{}.
\newblock \showarticletitle{Surveillance Capitalism and the Challenge of
  Collective Action}.
\newblock \bibinfo{journal}{\emph{New Labor Forum}} \bibinfo{volume}{28},
  \bibinfo{number}{1} (\bibinfo{year}{2019}), \bibinfo{pages}{10--29}.
\newblock
\urldef\tempurl%
\url{https://doi.org/10.1177/1095796018819461}
\showDOI{\tempurl}


\end{thebibliography}

{\footnotesize\noindent\vspace{1em} All URLs were last accessed on May 9th, 2024.}

\end{document}